\documentclass[12pt,a4paper]{article}
\usepackage[latin1]{inputenc}
\usepackage[english]{babel}
\usepackage[T1]{fontenc}
\usepackage{ae,aecompl}
\usepackage{amssymb,amsmath,amsfonts}
\usepackage{ifpdf}
\ifpdf
  \usepackage[pdftex,a4paper,hmargin=25mm,vmargin=24mm]{geometry}
  \usepackage[pdftex,bookmarks=true]{hyperref}
\else
  \usepackage[dvips,a4paper,hmargin=25mm,vmargin=24mm]{geometry}
  \usepackage[dvips,bookmarks=true]{hyperref}
\fi
\usepackage{enumerate}

\numberwithin{equation}{section}

\newcommand{\mc}[1]{\mathcal{#1}}
\newcommand{\bb}[1]{\mathbb{#1}}
\newcommand{\rd}{\mathrm{d}} 
\newcommand{\dx}[1]{\rd x^{#1}}
\newcommand{\e}[1]{\frac{\partial}{\partial x^{#1}}}
\newcommand{\X}[2]{\frac{\partial #1}{\partial #2}} 

\newcommand{\Aab}{A^{a_1 \cdots a_k}_{\phantom{a_1 \cdots a_k} b_1 \cdots b_l}} 
\newcommand{\Aaib}[1]{A^{a_1 \cdots a_{i-1} #1 a_{i+1} \cdots a_k}_{\phantom{a_1 \cdots a_{i-1} #1 a_{i+1} \cdots a_k} b_1 \cdots b_l}} 
\newcommand{\Aabi}[1]{A^{a_1 \cdots a_k}_{\phantom{a_1 \cdots a_k} b_1 \cdots b_{i-1} #1 b_{i+1} \cdots b_l}} 
\newcommand{\Aabc}{A^{a_1 \cdots a_k}_{\phantom{a_1 \cdots a_k} b_1 \cdots b_l c_1 \cdots c_p}} 
\newcommand{\A}{A^{\mu_1 \cdots \mu_k}_{\phantom{\mu_1 \cdots \mu_k} \nu_1 \cdots \nu_l}} 
\newcommand{\Amui}[1]{A^{\mu_1 \cdots \mu_{i-1} #1 \mu_{i+1} \cdots \mu_k}_{\phantom{\mu_1 \cdots \mu_{i-1} #1 \mu_{i+1} \cdots \mu_k} \nu_1 \cdots \nu_l}} 
\newcommand{\Anui}[1]{A^{\mu_1 \cdots \mu_k}_{\phantom{\mu_1 \cdots \mu_k} \nu_1 \cdots \nu_{i-1} #1 \nu_{i+1} \cdots \nu_l}} 
\newcommand{\Aform}{A^{\mu_1 \cdots \mu_k}_{\phantom{\mu_1 \cdots \mu_k} \nu_1 \cdots \nu_l \rho_1 \cdots \rho_p}} 
\newcommand{\B}{B^{\rho_1 \cdots \rho_m}_{\phantom{\rho_1 \cdots \rho_m} \sigma_1 \cdots \sigma_n}} 
\newcommand{\Brhoi}[1]{B^{\rho_1 \cdots \rho_{i-1} #1 \rho_{i+1} \cdots \rho_m}_{\phantom{\rho_1 \cdots \rho_{i-1} #1 \rho_{i+1} \cdots \rho_m} \sigma_1 \cdots \sigma_n}} 
\newcommand{\Bsigmai}[1]{B^{\rho_1 \cdots \rho_m}_{\phantom{\rho_1 \cdots \rho_m} \sigma_1 \cdots \sigma_{i-1} #1 \sigma_{i+1} \cdots \sigma_n}} 
\newcommand{\C}{C^{\lambda_1 \cdots \lambda_i}_{\phantom{\lambda_1 \cdots \lambda_j} \tau_1 \cdots \tau_j}} 
\newcommand{\Clambdai}[1]{C^{\lambda_1 \cdots \lambda_{i-1} #1 \lambda_{i+1} \cdots \lambda_i}_{\phantom{\lambda_1 \cdots \lambda_{i-1} #1 \lambda_{i+1} \cdots \lambda_j} \tau_1 \cdots \tau_j}} 
\newcommand{\Ctaui}[1]{C^{\lambda_1 \cdots \lambda_i}_{\phantom{\lambda_1 \cdots \lambda_j} \tau_1 \cdots \tau_{i-1} #1 \tau_{i+1} \cdots \tau_j}} 

\newcommand{\ALiea}{A^{\mu_1 \cdots \mu_k \phantom{\nu_1 \cdots \nu_l} a}_{\phantom{\mu_1 \cdots \mu_k} \nu_1 \cdots \nu_l}} 
\newcommand{\BLieb}{B^{\rho_1 \cdots \rho_m \phantom{\sigma_1 \cdots \sigma_n} b}_{\phantom{\rho_1 \cdots \rho_m} \sigma_1 \cdots \sigma_n}} 

\begin{document}
\begin{center}
{\Large Covariant star product on symplectic and Poisson\\
\vspace{.3em}
spacetime manifolds}\\
\vspace{1em}
M. Chaichian$^{a,b}$, M. Oksanen$^a$, A. Tureanu$^{a,b}$ and G. Zet$^c$\\
\vspace{1em}
\textit{${}^a$Department of Physics, University of Helsinki, P.O. Box 64,
FI-00014 Helsinki, Finland\\
${}^b$Helsinki Institute of Physics, P.O. Box 64, FI-00014 Helsinki, Finland\\
${}^c$Department of Physics, "Gh. Asachi" Technical University,\\Bd.
D. Mangeron 67, 700050 Iasi, Romania}
\end{center}

\begin{abstract}
A covariant Poisson bracket and an associated covariant star product in the sense of deformation quantization are defined on the algebra of tensor-valued differential forms on a symplectic manifold, as a generalization of similar structures that were recently defined on the algebra of (scalar-valued) differential forms. A covariant star product of arbitrary smooth tensor fields is obtained as a special case. Finally, we study covariant star products on a more general Poisson manifold with a linear connection, first for smooth functions and then for smooth tensor fields of any type. Some observations on possible applications of the covariant star products to gravity and gauge theory are made.
\end{abstract}

\section{Introduction}
Due to several convincing arguments arising from the quantum theory and the Einstein's theory of gravity, it is generally believed that the manifold structure of spacetime does not exist at distances equal and shorter than the Planck length and that the correct description of spacetime should be somehow noncommutative. Field theories defined on noncommutative spacetimes have been extensively studied during the last decades (for some reviews see \cite{douglas+nekrasov:2001,szabo:2003}). The canonically noncommutative spacetime structure, generated by the coordinate commutation relations
\begin{equation}
\left[ \hat{x}^\mu, \hat{x}^\nu \right] = i\theta^{\mu\nu}
\end{equation}
with a constant antisymmetric $\theta^{\mu\nu}$, and its Moyal star product have received most attention. Also the Lie algebraic structure, the quantum space structures and the symplectic and Poisson manifolds have been considered as possible descriptions of noncommutative spacetime. We consider the last two cases where the $\theta^{\mu\nu}(\hat{x})$ is a generally $\hat{x}$-dependent bivector field.

The main effects of the noncommutativity of spacetime on the theories of particle physics, most notably the Standard Model, have been extensively studied and by now some of their features are well understood. Understanding gravity on noncommutative spacetimes has proven to be a challenging effort. This is due to the difficulty to accommodate both the gravitational and the noncommutative structures of spacetime --- the classical geometrical large-distance structure and the noncommutativity of coordinates at short distances.

One of the standing issues of noncommutative gravity is the general covariance of the star product under spacetime diffeomorphisms. The diffeomorphism-covariance of a star product can be achieved in many ways. One way is to construct a star product that is by definition covariant under conventional spacetime diffeomorphism. This is the approach we will consider in this work. More specifically we consider spacetime as a symplectic manifold --- later as a more general Poisson manifold --- and seek to quantize such a spacetime by introducing a (noncommutative) covariant star product. This is done in the light of two recent approaches \cite{mccurdy+zumino:2009,tagliaferro:2008,vassilevich:2009} to the quantization of a symplectic spacetime manifold. We construct a diffeomorphism-covariant Poisson bracket and an associated star product of tensor-valued differential forms on such spacetime. A covariant star product of tensor fields is obtained as the special case of tensor-valued zero-forms. Possible applications of the obtained covariant star product to gravity and gauge theory are discussed.

Deformation quantization of more general Poisson manifolds with a torsion-free linear connection has also been studied recently \cite{ammar+chloup+gutt:2008} and a universal covariant star product of functions has been constructed. We define a covariant Poisson bracket on a smooth manifold with a linear connection and propose an associated covariant star product of tensor fields on the Poisson manifold. The constraints that the connection is imposed to satisfy by these structures are studied. The possibility to relax the torsion-freeness condition of \cite{ammar+chloup+gutt:2008} in the case of a star product of functions is also considered.

For a recent review of deformation quantization see \cite{dito+stemheimer:2002}.

\section{On covariant derivative of tensors and differential forms}

The intent of this section is to review the concepts of connection and covariant derivative on smooth manifolds, providing some of the definitions and results that are used in the following sections, and to discuss some misunderstandings found in recent literature regarding these things.

\subsection{Connections and covariant derivatives}
We consider a smooth manifold $M$ and a linear connection on the tensor bundle $T(M)$ of $M$ and the associated covariant derivative.\footnote{We could equally well talk about an affine connection instead of a linear connection. See \cite{book:kobayashi+nomizu:1963}, Chapter~3, Theorem~3.3, for their relation.} The linear connection is given by a covariant derivative $\nabla$ that is a linear map
\begin{equation}
\nabla : T^{k, l}(M ) \rightarrow T^{k, l+1}(M) \, ,\label{nabla}
\end{equation}
where $T^{k, l}(M)$ is the vector space of smooth tensor fields of type $(k, l)$ on $M$, i.e. the space of smooth sections of the tensor product bundle $\otimes^k TM \otimes^l T^*M$
\begin{equation}
T^{k, l}(M) = \Gamma (\otimes^k TM \otimes^l T^*M) \, ,
\end{equation}
where $TM$ and $T^*M$ are the tangent bundle of $M$ and the cotangent bundle of $M$, respectively, $\otimes^k TM$ denotes the $k$-th tensor power of $TM$ and $\Gamma$ denotes the space of all smooth sections of the argument fiber bundle. We shall denote the algebra of tensor fields on $M$ by
\begin{equation}
\mc{T}(M) = \bigoplus_{k,l=0}^\infty T^{k, l}(M) \, .
\end{equation}
The covariant derivative $\nabla_X$ along a vector field $X \in \mc{X}(M) = \Gamma (TM)$ is a linear derivation that preserves the type of tensors
\begin{equation}
\nabla_X : T^{k, l}(M ) \rightarrow T^{k, l}(M) \label{nabla_X}
\end{equation}
and it is related to the connection \eqref{nabla} by
\begin{equation}
(\nabla_X A)(\alpha_1, \ldots, \alpha_k, X_1, \ldots, X_l) = (\nabla A)(X; \alpha_1, \ldots, \alpha_k, X_1, \ldots, X_l) \, ,\label{nabla_X_A_def}
\end{equation}
where the vector field $X$ in the covariant derivative $\nabla_X A$ of $A \in T^{k, l}(M)$ takes the place of the additional vector argument in $\nabla A \in T^{k, l+1}(M)$ provided by \eqref{nabla}  (see \cite{book:kobayashi+nomizu:1963}, Chapter~3, Section~2).\footnote{The additional argument $X$ in the \eqref{nabla_X_A_def} is the first one, because we want to have the arguments of $\nabla A$ in the same order as the corresponding tensor indices in the component notation $\nabla_\rho \A$.} This together with the requirements that $\nabla_X$ commutes with all contractions and acts on functions as the vector $X$ (directional derivative)
\begin{equation}
\nabla_X f = X(f) \, ,\ f \in \mc{F}(M) = \Gamma (M \times \bb{R}) \label{nabla_X_f}
\end{equation}
ensures that $\nabla$ satisfies the properties of a covariant differentation on $\mc{T}(M)$.\footnote{The linearity of a tensor $\nabla A$ in its arguments guarantees that $\nabla_{fX} = f \nabla_X$ and $\nabla_{X+Y} = \nabla_X + \nabla_Y$, for arbitrary  $f \in \mc{F}(M)$ and $X, Y \in \mc{X}(M)$.} The covariant derivative \eqref{nabla_X_A_def} can be written
\begin{equation}
\begin{split}
(\nabla_X A)(\alpha_1, \ldots, \alpha_k, X_1, \ldots, X_l) &= \nabla_X \bigl( A(\alpha_1, \ldots, \alpha_k, X_1, \ldots, X_l) \bigl) \\
&- \sum_{i=1}^k A(\alpha_1, \ldots, \nabla_X \alpha_i, \ldots, \alpha_k, X_1, \ldots, X_l) \\
&- \sum_{i=1}^l A(\alpha_1, \ldots, \alpha_k, X_1, \ldots, \nabla_X X_i, \ldots, X_l) \, ,\label{nabla_X_A}
\end{split}
\end{equation}
which follows from $\nabla_X$ being a derivation that commutes with all contractions (see \cite{book:kobayashi+nomizu:1963}, Chapter~3, Proposition~2.10). Thus the second covariant derivative of $A \in T^{k, l}(M)$ is
\begin{equation}
(\nabla^2 A) (X; Y; ) = \nabla_X (\nabla_Y A) - \nabla_{\nabla_X Y} A \, ,\label{nabla2_A}
\end{equation}
where each term is in $T^{k, l}(M)$ (see \cite{book:kobayashi+nomizu:1963}, Chapter~3, Proposition~2.12). The $n$-th covariant derivative can be obtained inductively.

\paragraph{Differential forms}
The vector space of differential forms of degree $p$ on $M$ is the space of smooth sections of the $p$-th exterior power of the cotangent bundle,
\begin{equation}
\Omega^p(M) = \Gamma (\wedge^p T^*M) \, .
\end{equation}
The algebra of differential forms on $M$ --- with the exterior product $\wedge$ as multiplication --- is the direct sum of the spaces of $p$-forms of all degrees $p$ and it shall be denoted by
\begin{equation}
\Omega(M) = \bigoplus_{p=0}^{\dim M} \Omega^p(M) \, .
\end{equation}
The covariant derivative of a differential form on $M$ is defined similarly as for any other tensor field on $M$ (see above). However, the algebra $\Omega(M)$ is not closed under a covariant differentation $\nabla$. For example restricting the domain of $\nabla$ to $\Omega^p(M)$ we have
\begin{equation}
\nabla : \Gamma (\wedge^p T^*M) \rightarrow \Gamma (T^*M \otimes \wedge^p T^*M) \, ,\label{nabla_on_OmegaM}
\end{equation}
where the range is the space of covector-valued $p$-forms. Thus we have to consider tensor-valued differential forms.

The vector space of $(k, l)$-tensor-valued differential forms of degree $p$ shall be denoted by
\begin{equation}
\Omega^p(M, T^{k, l}) = \Gamma ( \otimes^k TM \otimes^l T^*M \otimes \wedge^p T^*M ) \, ,\label{tensor-valued_forms}
\end{equation}
where $T^{k, l}$ abbreviates the tensor product bundle $\otimes^k TM \otimes^l T^*M$.\footnote{We shall also refer to elements of $\Omega^p(M, T^{k, l})$ as $(k, l)$-tensor-valued $p$-forms.} Note that $\Omega^0(M, T^{k, l}) = T^{k, l}(M)$ and $\Omega^p(M, T^{0, 0}) = \Omega^p(M)$. The algebra of all tensor-valued differential forms is defined as
\begin{equation}
\Omega(M, T) = \bigoplus_{p=0}^{\dim M} \bigoplus_{k, l = 0}^\infty  \Omega^p(M, T^{k, l}) \, ,\label{algebra_tvf}
\end{equation}
with the multiplication given by the generalized exterior product
\begin{equation}
\wedge : \Omega^p(M, T^{k, l}) \times \Omega^q(M, T^{m, n}) \rightarrow \Omega^{p+q}(M, T^{k, l} \otimes T^{m, n}) = \Omega^{p+q}(M, T^{k+m, l+n}) \, ,\label{exterior_product_def}
\end{equation}
The covariant derivative $\nabla$ maps $(k, l)$-tensor-valued $p$-forms to $(k, l+1)$-tensor-valued $p$-forms
\begin{equation}
\nabla : \Omega^p(M, T^{k, l}) \rightarrow \Omega^p(M, T^{k, l+1}) \, .
\end{equation}
We also define an exterior covariant derivative $D$ that is the natural extension of the exterior derivative $\rd: \Omega^p(M) \rightarrow \Omega^{p+1}(M)$ and $\nabla$ on $\Omega(M, T)$. It maps tensorial $p$-forms to tensorial $(p+1)$-forms of the same type
\begin{equation}
D : \Omega^p(M, T^{k, l}) \rightarrow \Omega^{p+1}(M, T^{k, l}) \, ,\label{exterior_covariant_derivative}
\end{equation}
which we shall discuss more shortly (see also \cite{book:kobayashi+nomizu:1963}, Chapter~2, Section~5).

\paragraph{Local smooth frames, the connection one-form, the torsion and the curvature two-forms and the exterior covariant derivative}
A connection one-form $\omega^a_{\phantom{a}b}$ of $\nabla$ is associated to a local smooth frame $\{ e_a \}_{a=1}^{\dim M}$ of the tangent bundle $TM$ over an open set $U$ of $M$ over which $TM$ is trivial. It is defined by
\begin{equation}
 \nabla e_b = \omega^a_{\phantom{a}b} \otimes e_a \, .
\end{equation}
The connection $\nabla$ on $TM$ (restricted over $U$) is given by
\begin{equation}
\nabla \phi = ( \rd \phi^a + \omega^a_{\phantom{a}b} \phi^b ) \otimes e_a \, ,\label{nabla_phi}
\end{equation}
where $\phi = \phi^a e_a \in \Gamma(TM)$ over $U$ and $\rd$ is the exterior derivative. On the cotangent bundle $T^*M$, the dual bundle of $TM$, we can setup a local smooth frame $\{ e^a \}_{a=1}^{\dim M}$ over $U$ that is dual to the frame of $TM$, $\langle e^a, e_b \rangle = \delta^a_b$. Thus the connection on $T^*M$ over $U$ is given by $\nabla e^a = - \omega^a_{\phantom{a}b} \otimes e^b$ and
\begin{equation}
\nabla \psi = ( \rd \psi_b - \omega^a_{\phantom{a}b} \psi_a ) \otimes e^b \, ,\label{nabla_varphi}
\end{equation}
where $\psi = \psi_a e^a \in \Gamma(T^*M)$. Extension to the tensor bundle $T(M)$ is straightforward,\footnote{$\nabla$ has the standard Leibniz rule, $\nabla (A \otimes B) = \nabla A \otimes B + A \otimes \nabla B$, and similarly for the exterior product, $\nabla (A \wedge B) = \nabla A \wedge B + A \wedge \nabla B$.}
e.g. for $A = \Aab e_{a_1} \otimes \cdots \otimes e_{a_k} \otimes e^{b_1} \otimes \cdots \otimes e^{b_l} \in \Gamma( \otimes^k TM \otimes^l T^*M )$ over $U$ we have
\begin{multline}
\nabla A = \left( \rd \Aab + \sum_{i=1}^k \omega^{a_i}_{\phantom{a_i}c} \Aaib{c} - \sum_{i=1}^l \omega^c_{\phantom{c}b_i} \Aabi{c} \right) \\
\otimes e_{a_1} \otimes \cdots \otimes e_{a_k} \otimes e^{b_1} \otimes \cdots \otimes e^{b_l} \, .\label{nabla_Aab}
\end{multline}

All the other local smooth frames of $T^*M$ and $TM$ can be obtained through \emph{local} linear transformations
\begin{equation}
e^{'a} = \Lambda^a_{\phantom{a}b} e^b \, ,\quad e'_a = e_b (\Lambda^{-1})^b_{\phantom{b}a} \, ,\label{frame_transformation}
\end{equation}
where in the general case $\Lambda \in GL(T_p M) \cong GL(\dim M, \bb{R})$, but additional structures on $M$ can restrict the local symmetry group to a subgroup of $GL(\dim M, \bb{R})$. The components of tensor fields transform as
\begin{equation}
A^{'a_1 \cdots a_k}_{\phantom{'a_1 \cdots a_k} b_1 \cdots b_l} = \Lambda^{a_1}_{\phantom{a_1}c_1} \cdots \Lambda^{a_k}_{\phantom{a_k}c_k} A^{c_1 \cdots c_k}_{\phantom{c_1 \cdots c_k} d_1 \cdots d_l} (\Lambda^{-1})^{d_1}_{\phantom{d_1} b_1} \cdots (\Lambda^{-1})^{d_l}_{\phantom{d_l} b_l}
\label{tensor_transformation}
\end{equation}
and the connection one-form has the transformation rule
\begin{equation}
\omega^{'a}_{\phantom{'a}b} = \Lambda^a_{\phantom{a}c} \omega^c_{\phantom{c}d} (\Lambda^{-1})^d_{\phantom{d}b} - \rd \Lambda^a_{\phantom{a}c} (\Lambda^{-1})^c_{\phantom{c}b} \, .
\end{equation}

For tensor-valued differential forms we use notation where the tensor indices are visible and the antisymmetric form components are hidden, e.g. $A \in \Omega^p(M, T^{k, l})$ is written
\begin{equation}
\Aab = \frac{1}{p!} \Aabc e^{c_1} \wedge \cdots \wedge e^{c_p} \, .\label{A_tensor-valued_form}
\end{equation}
The torsion two-form $T^a$ and the curvature two-form $R^a_{\phantom{a}b}$ of the connection are defined by
\begin{align}
T^a &= D e^a = \rd e^a + \omega^a_{\phantom{a}b} \wedge e^b \, ,\label{torsion_two-form}\\
R^a_{\phantom{a}b} &= \rd \omega^a_{\phantom{a}b} + \omega^a_{\phantom{a}c} \wedge \omega^c_{\phantom{c}b} \, ,\label{curvature_two-form}
\end{align}
where $D$ is the exterior covariant derivative \eqref{exterior_covariant_derivative} that is defined for a tensor-valued differential form \eqref{A_tensor-valued_form} as the linear map
\begin{equation}
\begin{split}
D \Aab &= \rd \Aab + \sum_{i=1}^k \omega^{a_i}_{\phantom{a_i} c} \wedge \Aaib{c} \\
&- \sum_{i=1}^l \omega^c_{\phantom{c} b_i} \wedge \Aabi{c} \, .\label{DAab}
\end{split}
\end{equation}
Unlike the exterior derivative $\rd \Aab$ the exterior covariant derivative \eqref{DAab} has the correct tensor transformation rule \eqref{tensor_transformation} under local frame transformations \eqref{frame_transformation}. The second exterior covariant derivative consist of contractions with the curvature two-form \eqref{curvature_two-form}
\begin{equation}
D^2 \Aab = \sum_{i=1}^k R^{a_i}_{\phantom{a_i} c} \wedge \Aaib{c} -  \sum_{i=1}^l R^c_{\phantom{c} b_i} \wedge \Aabi{c} \, .
\end{equation}
Taking exterior covariant derivatives of \eqref{torsion_two-form} and \eqref{curvature_two-form} yields the Bianchi identities
\begin{align}
D T^a &= R^a_{\phantom{a} b} \wedge e^b \, ,\label{DTa}\\
D R^a_{\phantom{a} b} &= 0 \, .\label{DRab}
\end{align}

\paragraph{Local coordinates}
Introducing a local coordinate system $\{x^\mu\}_{\mu=1}^{\dim M}$ on the open set $U$ of $M$ enables us to use the full component notation of tensor calculus --- the formalism conventionally used in physics. It enables us to locally write the covariant derivative \eqref{nabla_Aab} of a tensor field $A \in T^{k, l}(M)$ along the basis vector $\e{\mu}$ as
\begin{equation}
\begin{split}
\nabla_\mu \Aab &= \partial_\mu \Aab + \sum_{i=1}^k \omega^{\phantom{\mu} a_i}_{\mu\phantom{a_i}c} \Aaib{c} \\
&- \sum_{i=1}^l \omega^{\phantom{\mu} c}_{\mu\phantom{c}b_i} \Aabi{c} \, ,\label{nabla_mu_Aab}
\end{split}
\end{equation}
where $\omega^{\phantom{\mu} a}_{\mu \phantom{a} b} \dx{\mu} = \omega^a_{\phantom{a}b}$, $\nabla_\mu e_b = \omega^{\phantom{\mu} a}_{\mu \phantom{a} b} e_a$ and $\nabla_\mu e^a = - \omega^{\phantom{\mu} a}_{\mu \phantom{a} b} e^b$. This is the local form of \eqref{nabla_X_A}.

Since the fibers of $TM$ and $T^*M$ over each $p \in M$ are the tangent space $T_p M$  and the cotangent space $T^*_p M$ of $M$ at $p$ respectively, the local frames of $TM$ and $T^*M$ over each $p \in M$ are smoothly related to the coordinate bases $\e{\mu}$ and $\dx{\mu}$ of $T_p M$ and $T^*_p M$ respectively through (orientation preserving) linear transformations
\begin{equation}
e_a = e_a^{\phantom{a}\mu} \e{\mu} \, ,\quad e^a = e^a_{\phantom{a}\mu} \dx{\mu} \, ,
\end{equation}
where $e_a^{\phantom{a}\mu}$ as a matrix is a $GL^+(\dim M, \bb{R})$-valued smooth function on $M$ and $e^a_{\phantom{a}\mu}$ is the inverse of $e_a^{\phantom{a}\mu}$; $e_a^{\phantom{a}\mu} e^b_{\phantom{b}\mu} = \delta_a^b$, $e_a^{\phantom{a}\mu} e^a_{\phantom{a}\nu} = \delta^\mu_\nu$.\footnote{$GL^+(\dim M, \bb{R}) = \{ g \in GL(\dim M, \bb{R}) : \det g > 0 \}$}
The functions $e_a^{\phantom{a}\mu}$ and $e^a_{\phantom{a}\mu}$ enable us to transform components of tensors between the coordinate and noncoordinate bases.

A $(k, l)$-tensor-valued $p$-form \eqref{A_tensor-valued_form} behaves as a $(k, l+p)$-tensor field under the covariant derivative \eqref{nabla_Aab}
\begin{equation}
\nabla_\mu \Aab = \frac{1}{p!} \left( \nabla_\mu \Aabc \right) e^{c_1} \wedge \cdots \wedge e^{c_p} \, ,\label{nabla_mu_Aabform}
\end{equation}
where the expression inside the parenthesis is given by \eqref{nabla_mu_Aab}.

\paragraph{Using a coordinate basis for $T(M)$}
We can even choose the local frames of $TM$ and $T^*M$ to coincide with a coordinate basis of tangent spaces, $e_a = \e{a}$, and cotangent spaces, $e^a = \dx{a}$. When this choice is made, we conventionally choose to work with one kind of indices, $a \rightarrow \mu$ etc., and rename the connection one-form $\omega^a_{\phantom{a}b} \rightarrow \Gamma^\rho_\nu$ and the connection coefficients $\omega^{\phantom{\mu} a}_{\mu \phantom{a} b} \rightarrow \Gamma^\rho_{\mu\nu}$.
The covariant derivative is now defined by
\begin{gather}
\nabla_\mu : T^{k, l}(M ) \rightarrow T^{k, l+1}(M) \, ,\label{nabla_mu} \\
\begin{split}
\nabla_\rho \A &= \partial_\mu \A + \sum_{i=1}^k \Gamma^{\mu_i}_{\rho\sigma} \Amui{\sigma} \\
&- \sum_{i=1}^l \Gamma^\sigma_{\rho\nu_i} \Anui{\sigma} \, .\label{nabla_rho_A}
\end{split}
\end{gather}

General coordinate transformations, $x \rightarrow x' = x'(x)$, are a specific class of frame transformations \eqref{frame_transformation} with the local transformation matrix
\begin{equation}
\Lambda^\mu_{\phantom{\mu}\nu} = \X{x^{'\mu}}{x^\nu} \, .
\end{equation}

\subsection{Criticism}
\label{subsec:Criticism}
It is important to understand that the algebra of differential forms $\Omega(M)$ is not closed under the covariant derivation $\nabla$ (equivalently under $\nabla_\mu$ in a coordinate basis). The covariant derivative $\nabla \omega$ of a $p$-form $\omega$ is a smooth section of the  product bundle $T^*M \otimes \wedge^p T^*M$. In other words $\nabla_\mu \omega$ is a $(0, 1)$-tensor-valued $p$-form. This is not acknowledged in \cite{tagliaferro:2008}, where the covariant derivative $\nabla_\mu \omega$ of a $p$-form $\omega$ along the basis vector $e_\mu$ is considered to be a $p$-form, which leads to some serious problems.

Differential forms are frame-independent objects that exist independent of \emph{any} coordinate system. $\nabla_\mu \omega$ is clearly a frame-dependent object that transforms as a component of a covector under general coordinate transformations.

The convention ``$\nabla_\mu$ acts nontrivially only on the bases $e_\mu$ and $\dx{\mu}$'' in \cite{tagliaferro:2008} is inconsistently executed. The property \eqref{nabla_X_A} is violated, when some of the contractions are differentiated with $\nabla_\mu$. As an example we consider the covariant derivative of the contraction of a bivector $\theta^{\mu\nu}$ and two covariant derivatives $\nabla_\mu \alpha$ and $\nabla_\nu \beta$ of differential forms $\alpha$ and $\beta$,
\begin{equation}
\nabla_\mu \left( \theta^{\nu\rho} \nabla_\nu \alpha \nabla_\rho \beta \right) = ( \nabla_\mu \theta^{\nu\rho} ) \nabla_\nu \alpha \nabla_\rho \beta + \theta^{\nu\rho} \left( \nabla_\mu \nabla_\nu \alpha \nabla_\rho \beta + \nabla_\nu \alpha \nabla_\mu \nabla_\rho \beta \right) \, .\label{contraction_example}
\end{equation}
Clearly we \emph{cannot} write $\nabla_\mu \theta^{\nu\rho} = \partial_\mu \theta^{\nu\rho}$, as is done in similar calculations of \cite{tagliaferro:2008} (see, \cite{tagliaferro:2008} Appendices~B.5 and C for these calculations), without trivializing the connection. The tensorial nature of $\tilde{R}^{\mu\nu}$ is correctly recognized in these calculations (see also the Appendix~A of \cite{tagliaferro:2008}), but the bivector $\theta^{\mu\nu}$ is treated as a function.

Moreover, in \cite{tagliaferro:2008} the second covariant derivatives $\nabla_\mu \nabla_\nu \alpha$ of a $p$-form $\alpha$ are incorrectly calculated, so that the commutator of second covariant derivatives of $\alpha$, 
\begin{equation}
[\nabla_\mu, \nabla_\nu] \alpha_{\rho_1 \cdots \rho_p} = - T^\sigma_{\phantom{\sigma}\mu\nu} \nabla_\sigma \alpha_{\rho_1 \cdots \rho_p}
 - \sum_{i=1}^p R^\sigma_{\phantom{\sigma} \rho_i \mu\nu} \alpha_{\rho_1 \cdots \rho_{i-1} \sigma \rho_{i+1} \cdots \rho_p} \, ,\label{cocd_component}
\end{equation}
contains only the curvature contributions, but not the torsion contribution.\footnote{If one wants to use the above mentioned convention for $\nabla_\mu$, one should calculate the second covariant derivative of $\alpha$ as $\nabla_\mu ( \dx{\nu} \otimes ( \nabla_\nu \alpha ) )$.}
This is an implication of the failure to fully recognize the additional argument vector provided by the covariant derivative.

Due to these problem in the covariant derivative of \cite{tagliaferro:2008}, the star product proposed in \cite{tagliaferro:2008} is neither truly associative nor covariant. The associativity property of the star product is found to be satisfied only because the covariant derivatives in the double Poisson brackets like $\{\{ \alpha, \beta\}, \gamma\}$ are calculated incorrectly.

These problems with the covariant derivative found in \cite{tagliaferro:2008} have been recently corrected in \cite{mccurdy+zumino:2009}, where the formalism of \cite{tagliaferro:2008} is reconsidered by using correct definitions. In \cite{mccurdy+zumino:2009} the covariant derivative $\nabla_\mu$ is correctly taken on tensor fields of any type and one does not try to extend the algebra of differential forms by the covariant derivatives.

\section{Generalization of the Poisson structure and the star product of differential forms to the algebra of tensor-valued differential forms on a symplectic manifold}
\label{sec:tensor-valued_forms}

\subsection{Poisson algebra of differential forms}
Consider the graded differential Poisson algebra of differential forms on a symplectic manifold $M$ studied in \cite{chu+ho:1997,tagliaferro:2008,mccurdy+zumino:2009,beggs+majid:2006}.

The Poisson bracket of functions $f, g \in \mc{F}(M)$ is defined by
\begin{equation}
\{ f, g \} = \theta( \rd f, \rd g ) = \theta^{\mu\nu} \partial_\mu f \partial_\nu g \, .\label{Pb}
\end{equation}
The Jacobi identity of the Poisson bracket requires that the Poisson bivector satisfies
\begin{equation}
\sum_{(\mu, \nu, \rho)} \theta^{\mu\sigma} \partial_\sigma \theta^{\nu\rho} = 0 \, ,\label{Jacobi}
\end{equation}
where the sum is over cyclic permutations. The Poisson bivector $\theta$ is assumed to be nondegenerate, so that it has an inverse $\omega$ that satisfies $\omega_{\mu\nu} \theta^{\nu\rho} = \delta_\mu^\rho$. It can be shown that \eqref{Jacobi} is equivalent to $\omega$ being a closed form, $\rd \omega = 0$ \cite{chu+ho:1997}. The closed nondegenerate two-form $\omega$ on $M$ is called the symplectic form.

The Poisson bracket of a function and a differential form $\alpha \in \Omega(M)$ (of degree one at first and then of any degree)
\begin{equation}
\{ f, \alpha \} =\nabla_{X_f} \alpha = \theta^{\mu\nu} \partial_\mu f \nabla_\nu \alpha  \label{Pb_function_and_form}
\end{equation}
is a covariant derivation of $\alpha$ and therefore defines a linear connection on $M$. By using the connection coefficients $\Gamma^\rho_{\mu\nu}$ we can define two connections $\nabla$ and $\tilde{\nabla}$ with the connection one-forms
\begin{equation}
\Gamma^\rho_\nu = \Gamma^\rho_{\mu\nu} \dx{\mu} \quad \text{and} \quad \tilde{\Gamma}^\rho_\mu = \Gamma^\rho_{\mu\nu} \dx{\nu} \label{connection_forms}
\end{equation}
respectively, which are different when the torsion \eqref{torsion_two-form},
\begin{equation}
T^\rho = \Gamma^\rho_\nu \wedge \dx{\nu} = \dx{\mu} \wedge \tilde{\Gamma}^\rho_\mu \, ,\label{torsion}
\end{equation}
does not vanish, $T^\rho_{\phantom{\rho}\mu\nu} = 2\Gamma^\rho_{[\mu\nu]} \neq 0$. The Leibniz rule of the Poisson bracket, $\rd \{ f, g \} = \{ \rd f, g \} + \{ f, \rd g \}$, implies that the connection $\tilde{\nabla}$ satisfies
\begin{equation}
\tilde{\nabla}_\mu \theta^{\nu\rho} = \partial_\mu \theta^{\nu\rho} + \Gamma^\nu_{\sigma\mu} \theta^{\sigma\rho} + \Gamma^\rho_{\sigma\mu} \theta^{\nu\sigma} = 0 \, ,\label{tildenabla_symplectic}
\end{equation}
i.e. $\tilde{\nabla}$ is a symplectic connection.\footnote{We call a connection $\nabla$ symplectic if $\omega_{\mu\nu}$ or equivalently $\theta^{\mu\nu}$ is covariantly constant under the covariant derivative.} Together \eqref{Jacobi} and \eqref{tildenabla_symplectic} imply two covariant versions of the Jacobi identity
\begin{equation}
\sum_{(\mu, \nu, \rho)} \theta^{\mu\sigma} \nabla_\sigma \theta^{\nu\rho} = 0
\quad \text{and} \quad
\sum_{(\mu, \nu, \rho)} \theta^{\mu\sigma} \theta^{\nu\lambda} T^\rho_{\phantom{\rho}\sigma\lambda} = 0 \, .\label{Jacobi_covariant}
\end{equation}
Imposing either $\nabla_\mu \theta^{\nu\rho} = 0$ or $T^\rho_{\phantom{\rho}\mu\nu} = 0$ would lead to a single torsion-free symplectic connection $\nabla = \tilde{\nabla}$, but this is not necessary. The curvature two-forms \eqref{curvature_two-form} of $\nabla$ and $\tilde{\nabla}$ are given by
\begin{equation}
R^\mu_{\phantom{\mu}\nu} = \rd \Gamma^\mu_\nu + \Gamma^\mu_\rho \wedge \Gamma^\rho_\nu
\quad \text{and} \quad
\tilde{R}^\mu_{\phantom{\mu}\nu} = \rd \tilde{\Gamma}^\mu_\nu + \tilde{\Gamma}^\mu_\rho \wedge \tilde{\Gamma}^\rho_\nu  \label{curvatures}
\end{equation}
respectively, and we use the Poisson bivector $\theta^{\mu\nu}$ to raise their lower index, e.g.
\begin{equation}
\tilde{R}^{\mu\nu} = \theta^{\mu\rho} \tilde{R}^\nu_{\phantom{\nu}\rho} \, .\label{tildeR}
\end{equation}
The curvature two-form of a symplectic connection $\tilde{\nabla}$ is symmetric $\tilde{R}^{\mu\nu} = \tilde{R}^{\nu\mu}$.\footnote{\eqref{tildenabla_symplectic} implies: $0 = [ \tilde{\nabla}_\rho, \tilde{\nabla}_\sigma ] \theta^{\mu\nu} = - T^\lambda_{\phantom{\lambda}\rho\sigma} \tilde{\nabla}_\lambda \theta^{\mu\nu} + \tilde{R}^\mu_{\phantom{\mu}\lambda\rho\sigma} \theta^{\lambda\nu} + \tilde{R}^\nu_{\phantom{\mu}\lambda\rho\sigma} \theta^{\mu\lambda} = - \tilde{R}^{\nu\mu}_{\phantom{\nu\mu}\rho\sigma} + \tilde{R}^{\mu\nu}_{\phantom{\mu\nu}\rho\sigma}$.}
Note that, unlike $\tilde{\nabla}_\mu$, the covariant derivative $\nabla_\mu$ does not commute with the raising of indices with $\theta^{\mu\nu}$, because $\nabla$ is not symplectic. Indeed \eqref{tildenabla_symplectic} implies
\begin{equation}
\nabla_\mu \theta^{\nu\rho} = T^\nu_{\phantom{\nu}\mu\sigma} \theta^{\sigma\rho} + T^\rho_{\phantom{\rho}\mu\sigma} \theta^{\nu\sigma} \, .\label{nabla_theta}
\end{equation}

The unique Poisson bracket of differential forms $\alpha, \beta \in \Omega(M)$ of nonzero degrees that is consistent with the graded differential Poisson algebra has been defined in \cite{tagliaferro:2008,mccurdy+zumino:2009}
\begin{equation}
\{ \alpha, \beta \} = \theta^{\mu\nu} \nabla_\mu \alpha \wedge \nabla_\nu \beta + (-1)^{\deg(\alpha)} \tilde{R}^{\mu\nu} \wedge i_\mu \alpha \wedge i_\nu \beta \, ,\label{Pb_form}
\end{equation}
where $\deg(\alpha)$ denotes the degree of $\alpha$ and $i_\mu \alpha$ is the interior product of $\alpha$ with the $\mu$-th basis vector. Covariant derivatives of contractions like \eqref{contraction_example}, including multiple covariant derivatives, are present when several Poisson brackets \eqref{Pb_form} are taken, e.g. $\{ \alpha, \{ \beta, \gamma \} \}$. In order for the Poisson bracket \eqref{Pb_form} to satisfy the graded Jacobi identity the connections have to satisfy the following additional constraints
\begin{gather}
R^\mu_{\phantom{\mu}\nu\rho\sigma} = 0 \, ,\label{R_zero} \\
\nabla_\lambda \tilde{R}^{\mu\nu}_{\phantom{\mu\nu}\rho\sigma} = 0 \, ,\label{tildeR_cov_constant}\\
\sum_{(\mu, \nu, \rho)} \tilde{R}^{\mu\sigma} \wedge i_\sigma \tilde{R}^{\nu\rho} = 0 \, ,\label{RiR}
\end{gather}
where the last constraint \eqref{RiR} is, however, implied by the two former constraints, the Leibniz rule and the Jacobi identity \eqref{Jacobi} \cite{beggs+majid:2006,mccurdy+zumino:2009}.\footnote{\eqref{R_zero} is implied by the Jacobi identity for two functions and one one-form and \eqref{tildeR_cov_constant} by the Jacobi identity for one function and two one.forms \cite{chu+ho:1997}.}

\subsection{Poisson algebra of tensor-valued differential forms}
We want to extend the graded differential Poisson algebra of differential forms by the covariant derivation $\nabla$. This is achieved by generalizing the Poisson bracket \eqref{Pb_form} for tensor-valued differential forms. In other words the Poisson bracket should be generalized to accept forms whose components have additional tensor indices. Then Poisson brackets like $\{ \nabla_\mu \alpha, \beta \}$ will be naturally defined. This would enable us to define the related star product for all tensor-valued differential forms, which enlarges the applicability of the formalism. The curvature two-forms $R^{\mu\nu}$ and $\tilde{R}^{\mu\nu}$ and the torsion two-form $T^\mu$ are examples of such forms. Such star product could indeed be useful for defining noncommutative deformations of gravitational theories, whose actions involve the curvature two-form(s). Next we propose such a formalism that generalizes the approach of \cite{tagliaferro:2008,mccurdy+zumino:2009}, and also corrects the misunderstandings found in \cite{tagliaferro:2008}.

\subsubsection*{The algebra of tensor-valued differential forms}
We choose to work in a local coordinate system $\{ x^\mu \}_{\mu=1}^{\dim M}$ of $M$. This approach can, however, be repeated by using any local smooth frame of $\Omega(M, T)$, with the frame transformations \eqref{frame_transformation} defined to be compatible with the symplectic structure of $M$.

The exterior product \eqref{exterior_product_def} of two tensor-valued differential forms $A \in \Omega^p(M, T^{k, l})$ and $B \in \Omega^q(M, T^{m, n})$,
\begin{align}
\A &= \frac{1}{p!} \Aform \dx{\rho_1} \wedge \cdots \wedge \dx{\rho_p} \, ,\\
B^{\mu_1 \cdots \mu_m}_{\phantom{\mu_1 \cdots \mu_m} \nu_1 \cdots \nu_n} &= \frac{1}{q!} B^{\mu_1 \cdots \mu_m}_{\phantom{\mu_1 \cdots \mu_m} \nu_1 \cdots \nu_n \rho_1 \cdots \rho_q} \dx{\rho_1} \wedge \cdots \wedge \dx{\rho_q} \, ,
\end{align}
is a tensor-valued differential form $A \wedge B \in \Omega^{p+q}(M, T^{k+m, l+n})$ defined by
\begin{multline}
(A \wedge B)^{\mu_1 \cdots \mu_{k+m}}_{\phantom{\mu_1 \cdots \mu_{k+m}} \nu_1 \cdots \nu_{l+n}} = \frac{1}{(p+q)!}  (A \wedge B)^{\mu_1 \cdots \mu_{k+m}}_{\phantom{\mu_1 \cdots \mu_{k+m}} \nu_1 \cdots \nu_{l+n} \rho_1 \cdots \rho_{p+q}} \dx{\rho_1} \wedge \cdots \wedge \dx{\rho_{p+q}} \\
= \frac{1}{p!q!} \Aform B^{\mu_{k+1} \cdots \mu_{k+m}}_{\phantom{\mu_{k+1} \cdots \mu_{k+m}} \nu_{l+1} \cdots \nu_{l+n} \rho_{p+1} \cdots \rho_{p+q}} \dx{\rho_1} \wedge \cdots \wedge \dx{\rho_{p+q}} \\
= \A \wedge B^{\mu_{k+1} \cdots \mu_{k+m}}_{\phantom{\mu_{k+1} \cdots \mu_{k+m}} \nu_{l+1} \cdots \nu_{l+n}} \, .\label{exterior_product}
\end{multline}
The exterior product \eqref{exterior_product} satisfies the following properties for arbitrary tensor-valued differential forms $A$, $B$ and $C$:
\begin{enumerate}
\item $A \wedge B = 0$ if $\deg(A) + \deg(B) > \dim(M)$.
\item Degree:
\begin{equation}
\deg(A \wedge B) = \deg(A) + \deg(B) \, .\label{wedge_degree}
\end{equation}
\item Symmetry:
\begin{equation}
\A \wedge \B = (-1)^{\deg(A) \deg(B)} \B \wedge \A \, .\label{wedge_symmetry}
\end{equation}
\item Associativity: $(A \wedge B) \wedge C = A \wedge (B \wedge C)$.
\end{enumerate}
It is necessary to write $\A$ instead of just $A$ in the exterior product \eqref{exterior_product} when the order of the factors is changeable as in \eqref{wedge_symmetry}, because the tensor product is generally noncommutative, $A \otimes B \neq B \otimes A$.\footnote{One can write \eqref{wedge_symmetry} equivalently as $A \wedge B = (-1)^{\deg(A) \deg(B)} (B \wedge A) \circ \sigma_{(k, l)}$, where the map $\sigma_{(k, l)}$ moves the first $k$ covector arguments and the first $l$ vector arguments over the rest of the arguments of each type, $\sigma_{(k, l)}(\alpha_1, \ldots, \alpha_{k+m}, X_1, \ldots, X_{l+n}) = (\alpha_{k+1}, \ldots, \alpha_{k+m}, \alpha_1, \ldots, \alpha_k, X_{l+1}, \ldots, X_{l+n}, X_1, \ldots, X_l)$. We, however, prefer to keep track of the order of the arguments with the tensorial indices.}

The interior product of tensor-valued differential forms can be defined so that it recognizes only the form part of tensor-valued differential forms. The interior product of $A \in \Omega^p(M, T^{k, l})$ with the coordinate basis vector $\e{\mu}$ is the map
\begin{equation}
\begin{split}
i_\mu &: \Omega^p(M, T^{k, l}) \rightarrow \Omega^{p-1}(M, T^{k, l+1}) \, ,\\
i_\rho \A &= \frac{1}{(p-1)!} A^{\mu_1 \cdots \mu_k}_{\phantom{\mu_1 \cdots \mu_k} \nu_1 \cdots \nu_l \rho \sigma_2 \cdots \sigma_p} \dx{\sigma_2} \wedge \cdots \wedge \dx{\sigma_p} \, .\label{interior_product}
\end{split}
\end{equation}
It satisfies
\begin{equation}
\begin{split}
i_\rho \left( \A \wedge \B \right) &= i_\rho \A \wedge \B \\
&+ (-1)^{\deg(A)} \A \wedge i_\rho \B
\end{split}
\label{interior_on_wedge}
\end{equation}
and $i_\mu i_\nu A = - i_\nu i_\mu A$. Zero-forms vanish under the interior product $i_\mu$.

The exterior covariant derivative $D$ \eqref{exterior_covariant_derivative} is used instead of the exterior derivative, because the latter maps tensorial differential forms to nontensorial ones. The exterior covariant derivative \eqref{DAab} is now written
\begin{equation}
\begin{split}
D \A &= \rd \A + \sum_{i=1}^k \Gamma^{\mu_i}_\rho \wedge \Amui{\rho} \\
&- \sum_{i=1}^l \Gamma^\rho_{\nu_i} \wedge \Anui{\rho} \, ,\label{DA}
\end{split}
\end{equation}
where the exterior derivative $\rd$ is given by
\begin{equation}
\rd \A = \frac{1}{p!} \partial_\sigma \Aform \dx{\sigma} \wedge \dx{\rho_1} \wedge \cdots \wedge \dx{\rho_p} \, .\label{exterior_derivative}
\end{equation}
$D$ satisfies the same Leibniz rule as $\rd$,
\begin{equation}
\begin{split}
D \left( \A \wedge \B \right) &= D \A \wedge \B \\
&+ (-1)^{\deg(A)} \A \wedge D \B \, .\label{Leibniz_D}
\end{split}
\end{equation}
The exterior covariant derivative $\tilde{D}$ of the other connection is defined analogously by using the connection one-form $\tilde{\Gamma}^\mu_\nu$ instead of $\Gamma^\mu_\nu$.

A connection also provides the covariant derivative on $\Omega(M, T)$
\begin{equation}
\nabla_\mu: \Omega^p(M, T^{k, l}) \rightarrow \Omega^p(M, T^{k, l+1})  \label{covariant_derivative}
\end{equation}
that is defined in \eqref{nabla_mu_Aab} and \eqref{nabla_mu_Aabform} (see also \eqref{nabla_mu}--\eqref{nabla_rho_A} for the present case of a coordinate basis).
The covariant derivative of a tensor-valued differential form can be written in a compact form as
\begin{equation}
\begin{split}
\nabla_\rho \A &= \partial_\rho \A + \sum_{i=1}^k \Gamma^{\mu_i}_{\rho\sigma} \Amui{\sigma} \\
&- \sum_{i=1}^l \Gamma^\sigma_{\rho \nu_i} \Anui{\sigma} - \tilde{\Gamma}^\sigma_\rho \wedge i_\sigma \A  \, ,\label{nabla_rho_Aform}
\end{split}
\end{equation}
where we denote
\begin{equation}
\partial_\sigma \A = \frac{1}{p!} \partial_\sigma \Aform \dx{\rho_1} \wedge \cdots \wedge \dx{\rho_p} \, .
\end{equation}
Defintion for the other covariant derivative $\tilde{\nabla}_\mu$ is analogous (replace $\Gamma^\rho_{\mu\nu}$ with $\Gamma^\rho_{\nu\mu}$ and $\tilde{\Gamma}^\rho_\mu$ with $\Gamma^\rho_\mu$). When the second covariant derivative $\nabla_\rho \nabla_\sigma \A$ is taken, the subscript $\sigma$ is treated as a covariant tensor index. The commutator of second covariant derivatives reads
\begin{equation}
\begin{split}
[\nabla_\rho, \nabla_\sigma] \A &= - T^\lambda_{\phantom{\lambda}\rho\sigma} \nabla_\lambda \A + \sum_{i=1}^k R^{\mu_i}_{\phantom{\mu_i}\lambda\rho\sigma} \Amui{\lambda} \\
&- \sum_{i=1}^l R^\lambda_{\phantom{\lambda}\nu_i \rho\sigma} \Anui{\lambda} - R^\lambda_{\phantom{\lambda}\tau\rho\sigma} \dx{\tau} \wedge i_\lambda \A  \, .\label{cocd_Aform}
\end{split}
\end{equation}
The covariant derivative has the Leibniz rule
\begin{equation}
\begin{split}
\nabla_\lambda \left( \A \wedge \B \right) &= \nabla_\lambda \A \wedge \B \\
&+ \A \wedge \nabla_\lambda \B \, .\label{Leibniz_cd}
\end{split}
\end{equation}

We can find a useful relation for $D$ and $\nabla_\rho$ by multiplying \eqref{nabla_rho_Aform} with $\dx{\rho} \wedge$ from left
\begin{equation}
\dx{\rho} \wedge \nabla_\rho \A = D \A - T^\rho \wedge i_\rho \A \, .
\end{equation}
We can even write it as a local operator identity
\begin{equation}
D = \dx{\mu} \wedge \nabla_\mu + T^\mu \wedge i_\mu \, .\label{D_and_nabla_relation}
\end{equation}
Once again a similar relation holds for the other connection
\begin{equation}
\tilde{D} = \dx{\mu} \wedge \tilde{\nabla}_\mu - T^\mu \wedge i_\mu \, .\label{tildeD_and_nabla_relation}
\end{equation}

We shall occasionally refer to both $\nabla$ and $D$ as the connection --- similarly for  $\tilde{\nabla}$ and $\tilde{D}$.

\subsubsection*{The Poisson bracket}
Now we can extend the Poisson bracket \eqref{Pb_form} for tensor-valued differential forms
\begin{equation}
\begin{split}
\left\{ \A, \B \right\} &= \theta^{\lambda\tau} \nabla_\lambda \A \wedge \nabla_\tau \B \\
&+ (-1)^{\deg(A)} \tilde{R}^{\lambda\tau} \wedge i_\lambda \A \wedge i_\tau \B \, .\label{Pb_tensor-valued}
\end{split}
\end{equation}

If either $\A$ or $\B$ (or both) is a tensor field of zero form degree, the Poisson bracket is defined by
\begin{equation}
\left\{ \A, \B \right\} = \theta^{\lambda\tau} \nabla_\lambda \A \nabla_\tau \B \, ,\label{Pb_tensor}
\end{equation}
which is also consistent with \eqref{Pb} and \eqref{Pb_function_and_form}.\footnote{We essentially consider that the interior product of a zero-form is zero.}

The Poisson bracket \eqref{Pb_tensor-valued} of tensor-valued differential forms satisfies the following properties of the graded differential Poisson algebra. For $A \in \Omega^p(M, T^{k, l})$ and $B \in \Omega^q(M, T^{m, n})$ and $C \in \Omega^r(M, T^{i, j})$ we have:
\begin{description}
\item[1. Bracket degree]
\begin{equation}
\deg\Bigl( \left\{ \A, \B \right\} \Bigr) = \deg(A) + \deg(B) \, ,
\end{equation}
is implied by the following properties. The covariant derivative \eqref{covariant_derivative} does not change the degree of tensor-valued differential forms, $\deg(\nabla_\mu A) = \deg(A)$. The interior product \eqref{interior_product} reduces the degree by one, $\deg(i_\mu A) = \deg(A) - 1$. The exterior product \eqref{exterior_product} has the degree \eqref{wedge_degree}.

\item[2. Graded symmetry]
\begin{equation}
\left\{ \A, \B \right\} = (-1)^{\deg(A)\deg(B)+1} \left\{ \B, \A \right\} \, ,
\end{equation}
follows from the symmetry property of the exterior product \eqref{wedge_symmetry} and from the antisymmetry of $\theta^{\mu\nu}$ and the symmetry of $\tilde{R}^{\mu\nu}$ under $\mu \leftrightarrow \nu$.

\item[3. Graded product]
\begin{multline}
\bigl\{ \A \wedge \B, \C \bigr\} \\
= \A \wedge \bigl\{ \B, \C \bigr\} \\
+ (-1)^{\deg(B) \deg(C)} \bigl\{ \A, \C \bigr\} \wedge \B \, ,\label{graded_product}
\end{multline}
follows from the Leibniz rule for $\nabla_\mu$ \eqref{Leibniz_cd} and the similar property for $i_\mu$ \eqref{interior_on_wedge} and the symmetry property of the exterior product \eqref{wedge_symmetry}.

\item[4. Leibniz rule]
\begin{equation}
\begin{split}
D \left\{ \A, \B \right\} &= \left\{ D \A, \B \right\} \\
&+ (-1)^{\deg(A)} \left\{ \A, D \B \right\} \, ,\label{Leibniz_D_on_Pb}
\end{split}
\end{equation}
By applying the Leibniz rule \eqref{Leibniz_D} to the left-hand side of \eqref{Leibniz_D_on_Pb} we obtain
\begin{multline}
D \left\{ \A, \B \right\} = D \theta^{\lambda\tau} \wedge \nabla_\lambda \A \wedge \nabla_\tau \B \\
+ \theta^{\lambda\tau} \Bigl( D \nabla_\lambda \A \wedge \nabla_\tau \B \\
+ (-1)^{\deg(A)} \nabla_\lambda \A \wedge D \nabla_\tau \B \Bigr) \\
+ (-1)^{\deg(A)} D \tilde{R}^{\lambda\tau} \wedge i_\lambda \A \wedge i_\tau \B \\
+ (-1)^{\deg(A)} \tilde{R}^{\lambda\tau} \wedge \Bigl( D i_\lambda \A \wedge i_\tau \B \\
+ (-1)^{\deg(A)-1} i_\lambda \A \wedge D i_\tau \B \Bigr) \, . \label{D_on_Pb}
\end{multline}
Then we use \eqref{D_and_nabla_relation} to calculate the relation of $D \nabla_\lambda \A$ and $\nabla_\lambda D \A$. First we calculate
\begin{equation}
D \nabla_\mu = \dx{\nu} \wedge \nabla_\nu \nabla_\mu + T^\nu \wedge i_\nu \nabla_\mu
\end{equation}
and
\begin{equation}
\nabla_\mu D = \dx{\nu} \wedge \nabla_\mu \nabla_\nu + \nabla_\mu T^\nu \wedge i_\nu + T^\nu \wedge \nabla_\mu i_\nu  \label{D_nabla}
\end{equation}
and find out that $\nabla_\mu$ and $i_\nu$ commute (follows from \eqref{interior_product} and \eqref{nabla_rho_Aform} by direct calculation, recalling $i_\nu i_\mu = - i_\mu i_\nu$)
\begin{equation}
i_\nu \nabla_\mu \A = \nabla_\mu i_\nu \A \, ,\label{i_and_nabla_commute}
\end{equation}
which then together imply
\begin{equation}
D \nabla_\mu = \nabla_\mu D + \dx{\nu} \wedge [\nabla_\nu, \nabla_\mu] - \nabla_\mu T^\nu \wedge i_\nu \, .
\end{equation}
Thus we obtain the relation
\begin{equation}
\begin{split}
D \nabla_\lambda \A &= \nabla_\lambda D \A + \dx{\rho} \wedge [\nabla_\rho, \nabla_\lambda] \A \\
&- \nabla_\lambda T^\rho \wedge i_\rho \A \, .\label{D_nabla_A}
\end{split}
\end{equation}
This result can as well be derived directly from the definitions \eqref{DA} and \eqref{nabla_rho_Aform}, but it is a lengthy calculation. By using the definitions \eqref{DA} and \eqref{interior_product} we obtain the relation of $D i_\lambda \A$ and $i_\lambda D \A$,
\begin{equation}
( D i_\lambda + i_\lambda D ) \A = \nabla_\lambda \A + i_\lambda T^\rho \wedge i_\rho \A \, ,\label{D_i_lambda}
\end{equation}
where we have also used \eqref{exterior_derivative}, \eqref{nabla_rho_Aform} and $i_\lambda T^\rho = \tilde{\Gamma}^\rho_\lambda - \Gamma^\rho_\lambda$.
Introducing the results \eqref{D_nabla_A} and \eqref{D_i_lambda} into \eqref{D_on_Pb} yields
\begin{multline}
D \left\{ \A, \B \right\} = \left\{ D \A, \B \right\} \\
+ (-1)^{\deg(A)} \left\{ \A, D \B \right\} \\
+ D \theta^{\lambda\tau} \wedge \nabla_\lambda \A \wedge \nabla_\tau \B \\
+ (-1)^{\deg(A)} \left( D \tilde{R}^{\lambda\tau} + \tilde{R}^{\phi\tau} \wedge i_\phi T^\lambda + \tilde{R}^{\lambda\phi} \wedge i_\phi T^\tau \right) \wedge i_\lambda \A \wedge i_\tau \B \\
+ \theta^{\phi\tau} \left( \tilde{R}^\lambda_{\phantom{\lambda}\phi} - \nabla_\phi T^\lambda \right) \wedge i_\lambda \A \wedge \nabla_\tau \B \\
+ (-1)^{\deg(A)} \theta^{\lambda\phi} \left( \tilde{R}^\tau_{\phantom{\tau}\phi} - \nabla_\phi T^\tau \right) \wedge \nabla_\lambda \A \wedge i_\tau \B \\
+ \theta^{\lambda\tau} \dx{\phi } \wedge \Bigl( [\nabla_\phi, \nabla_\lambda] \A \wedge \nabla_\tau \B \\
+ \nabla_\lambda \A \wedge [\nabla_\phi, \nabla_\tau] \B \Bigr) \, ,\label{D_on_Pb_unfinished}
\end{multline}
where some regrouping and simplifications have been done. For further simplification we calculate
\begin{equation}
\begin{split}
D \tilde{R}^{\mu\nu} + \tilde{R}^{\rho\nu} \wedge i_\rho T^\mu + \tilde{R}^{\mu\rho} \wedge i_\rho T^\nu &= \rd \tilde{R}^{\mu\nu} + \tilde{\Gamma}^\mu_\rho \wedge \tilde{R}^{\rho\nu} + \tilde{\Gamma}^\nu_\rho \wedge \tilde{R}^{\mu\rho} \\
&= \tilde{D} \tilde{R}^{\mu\nu} = \tilde{D} \left( \theta^{\mu\rho} \tilde{R}^\nu_{\phantom{\nu}\rho} \right) \\
&= \tilde{D} \theta^{\mu\rho} \wedge \tilde{R}^\nu_{\phantom{\nu}\rho} \, ,
\end{split}
\end{equation}
where \eqref{DRab} has been used in the last equality. As a final step we introduce \eqref{cocd_Aform} into the right-hand side of \eqref{D_on_Pb_unfinished} and combine the contributions of the first and the last term of \eqref{cocd_Aform} to the third, fifth and sixth term of  \eqref{D_on_Pb_unfinished}. In the third term of the resulting expression we calculate
\begin{equation}
D \theta^{\mu\nu} + \theta^{\rho\nu} i_\rho T^\mu + \theta^{\mu\rho} i_\rho T^\nu = \tilde{D} \theta^{\mu\nu} \, .
\end{equation}
Thus we obtain the result
\begin{multline}
D \left\{ \A, \B \right\} = \left\{ D \A, \B \right\} \\
+ (-1)^{\deg(A)} \left\{ \A, D \B \right\} \\
+ \tilde{D} \theta^{\lambda\tau} \wedge \nabla_\lambda \A \wedge \nabla_\tau \B \\
+ (-1)^{\deg(A)} \tilde{D} \theta^{\lambda\phi} \wedge \tilde{R}^\tau_{\phantom{\tau}\phi} \wedge i_\lambda \A \wedge i_\tau \B \\
+ \theta^{\phi\tau} \left( \tilde{R}^\lambda_{\phantom{\lambda}\phi} - \nabla_\phi T^\lambda + i_\phi R^\lambda_{\phantom{\lambda}\chi} \wedge \dx{\chi} \right) \wedge i_\lambda \A \wedge \nabla_\tau \B \\
+ (-1)^{\deg(A)} \theta^{\lambda\phi} \left( \tilde{R}^\tau_{\phantom{\tau}\phi} - \nabla_\phi T^\tau + i_\phi R^\tau_{\phantom{\tau}\chi} \wedge \dx{\chi} \right) \wedge \nabla_\lambda \A \wedge i_\tau \B \\
- \theta^{\lambda\tau} \Biggl[ \left( \sum_{i=1}^k i_\lambda R^{\mu_i}_{\phantom{\mu_i}\phi} \wedge \Amui{\phi} - \sum_{i=1}^l i_\lambda R^\phi_{\phantom{\phi}\nu_i} \wedge \Anui{\phi} \right) \wedge \\
\wedge \nabla_\tau \B + \nabla_\lambda \A \wedge \\
\wedge \left( \sum_{i=1}^m i_\tau R^{\mu_i}_{\phantom{\mu_i}\phi} \wedge \Brhoi{\phi} - \sum_{i=1}^n i_\tau R^\phi_{\phantom{\phi}\nu_i} \wedge \Bsigmai{\phi} \right) \Biggr] \, .\label{D_on_Pb_final}
\end{multline}
Hence for the Leibniz rule \eqref{Leibniz_D_on_Pb} to hold for arbitrary tensor-valued differential forms, we have to introduce the following constraints:
\begin{enumerate}
\item The connection $\tilde{D}$ is symplectic
\begin{equation}
\tilde{D} \theta^{\mu\nu} = \left( \tilde{\nabla}_\rho \theta^{\mu\nu} \right) \dx{\rho}= 0 \, .\label{tildeD_symplectic}
\end{equation}
\item The interior product of the curvature of $\nabla$ vanishes
\begin{equation}
i_\mu R^\nu_{\phantom{\nu}\rho} = 0 \, .\label{iR_zero}
\end{equation}
This implies that the curvature of $\nabla$ has to vanish, $R^\nu_{\phantom{\nu}\rho} = \frac{1}{2} \dx{\mu} \wedge i_\mu R^\nu_{\phantom{\nu}\rho} = 0$.
\item The curvature two-forms and the torsion two-form satisfy
\begin{equation}
\tilde{R}^\mu_{\phantom{\mu}\nu} - \nabla_\nu T^\mu + i_\nu R^\mu_{\phantom{\mu}\rho} \wedge \dx{\rho} = 0 \, .
\end{equation}
Taking \eqref{iR_zero} into account we obtain
\begin{equation}
\tilde{R}^\mu_{\phantom{\mu}\nu} = \nabla_\nu T^\mu \, .\label{R_T_constraint}
\end{equation}
\end{enumerate}

It would be quite tempting to require that the other connection $\tilde{D}$ satisfies a similar Leibniz rule as \eqref{Leibniz_D_on_Pb}. Such property would impose additional constraints on the connections and further restrict the geometry. However, we do not require such property for $\tilde{D}$, because the Poisson bracket \eqref{Pb_tensor-valued} has been defined with $\nabla$, not with $\tilde{\nabla}$, which makes $D$ the natural choice for the Leibniz rule \eqref{Leibniz_D_on_Pb}.

\item[5. Graded Jacobi identity]
\begin{equation}
\begin{split}
&\left\{ \A, \bigl\{ \B, \C \bigr\} \right\} \\
&+ (-1)^{\deg(A) [ \deg(B)+\deg(C) ]} \left\{ \B, \bigl\{ \C, \A \bigr\} \right\} \\
&+ (-1)^{[ \deg(A)+\deg(B) ] \deg(C)} \left\{ \C, \bigl\{ \A, \B \bigr\} \right\} = 0 \, ,\label{Jacobi_graded}
\end{split}
\end{equation}
First we calculate the Poisson bracket
\begin{multline}
\left\{ \A, \bigl\{ \B, \C \bigr\} \right\}  \\
= \theta^{\phi_1 \chi_1} \nabla_{\chi_1} \theta^{\phi_2 \chi_2} \nabla_{\phi_1} \A \wedge \nabla_{\phi_2} \B \wedge  \nabla_{\chi_2} \C \\
+ \theta^{\phi_1 \chi_1} \theta^{\phi_2 \chi_2} \Bigl( \nabla_{\phi_1} \A \wedge \nabla_{\chi_1} \nabla_{\phi_2} \B \wedge  \nabla_{\chi_2} \C \\
+ \nabla_{\phi_1} \A \wedge \nabla_{\phi_2} \B \wedge  \nabla_{\chi_1} \nabla_{\chi_2} \C \Bigr) \\
+ (-1)^{\deg(B)} \theta^{\phi_1 \chi_1}  \nabla_{\chi_1} \tilde{R}^{\phi_2 \chi_2} \wedge \nabla_{\phi_1} \A \wedge i_{\phi_2} \B \wedge  i_{\chi_2} \C \\
+ \theta^{\phi_1 \chi_1} \tilde{R}^{\phi_2 \chi_2} \wedge \Bigl( (-1)^{\deg(B)} \nabla_{\phi_1} \A \wedge i_{\phi_2} \nabla_{\chi_1} \B \wedge  i_{\chi_2} \C \\
+ (-1)^{\deg(B)} \nabla_{\phi_1} \A \wedge i_{\phi_2} \B \wedge  i_{\chi_2} \nabla_{\chi_1} \C  \\
+ (-1)^{\deg(A)} i_{\phi_2} \A \wedge i_{\chi_2} \nabla_{\phi_1} \B \wedge  \nabla_{\chi_1} \C \\
+ (-1)^{\deg(A)+\deg(B)} i_{\phi_2} \A \wedge \nabla_{\phi_1} \B \wedge  i_{\chi_2} \nabla_{\chi_1} \C \Bigr) \\
+ (-1)^{\deg(B)-1} \tilde{R}^{\phi_1 \chi_1} \wedge  i_{\chi_1} \tilde{R}^{\phi_2 \chi_2} \wedge i_{\phi_1} \A \wedge i_{\phi_2} \B \wedge  i_{\chi_2} \C \\
+ (-1)^{\deg(A)+\deg(B)} \tilde{R}^{\phi_1 \chi_1} \wedge \tilde{R}^{\phi_2 \chi_2} \wedge \Big( i_{\phi_1} \A \wedge i_{\chi_1} i_{\phi_2} \B \wedge  i_{\chi_2} \C  \\
+ (-1)^{\deg(B)+1} i_{\phi_1} \A \wedge i_{\phi_2} \B \wedge  i_{\chi_1} i_{\chi_2} \C \Bigr)  \, ,\label{Pb_ABC}
\end{multline}
where we have used \eqref{i_and_nabla_commute}. Cycling through $\A$, $\B$ and $\C$, using the symmetry property \eqref{wedge_symmetry} of the exterior product and introducing the expression \eqref{cocd_Aform} for the commutators of covariant derivatives, gives the left-hand side of the graded Jacobi identity \eqref{Jacobi_graded} as
\begin{multline}
\left\{ \A, \bigl\{ \B, \C \bigr\} \right\} \\
+ (-1)^{\deg(A) [ \deg(B)+\deg(C) ]} \left\{ \B, \bigl\{ \C, \A \bigr\} \right\} \\
+ (-1)^{[ \deg(A)+\deg(B) ] \deg(C)} \left\{ \C, \bigl\{ \A, \B \bigr\} \right\} \\
= \biggl[ \theta^{\phi_1 \chi_1} \left( \nabla_{\chi_1} \theta^{\phi_2 \chi_2} - \theta^{\phi_2 \psi} T^{\chi_2}_{\phantom{\chi_2}\chi_1 \psi} \right) + \theta^{\phi_2 \chi_1} \left( \nabla_{\chi_1} \theta^{\chi_2 \phi_1} - \theta^{\chi_2 \psi} T^{\phi_1}_{\phantom{\phi_1}\chi_1 \psi} \right) \\
+ \theta^{\chi_2 \chi_1} \left( \nabla_{\chi_1} \theta^{\phi_1 \phi_2} - \theta^{\phi_1 \psi} T^{\psi_2}_{\phantom{\psi_2}\chi_1 \psi} \right) \biggr] \nabla_{\phi_1} \A \wedge \nabla_{\phi_2} \B \wedge  \nabla_{\phi_3} \C \\
+ \theta^{\phi_1 \chi_1} \theta^{\phi_2 \chi_2} \Biggl[ \left( \sum_{i=1}^k R^{\mu_i}_{\phantom{\mu_i}\psi\phi_1 \phi_2} \Amui{\psi} - \sum_{i=1}^l R^\psi_{\phantom{\psi}\nu_i \phi_1 \phi_2} \Anui{\psi} \right) \wedge \\
\wedge \nabla_{\chi_1} \B \wedge  \nabla_{\chi_2} \C + \nabla_{\phi_1} \A \wedge \\
\wedge \left( \sum_{i=1}^k R^{\rho_i}_{\phantom{\rho_i}\psi\chi_1 \phi_2} \Brhoi{\psi} - \sum_{i=1}^l R^\psi_{\phantom{\psi}\sigma_i \chi_1 \phi_2} \Bsigmai{\psi} \right) \wedge \\
\wedge  \nabla_{\chi_2} \C + \nabla_{\phi_1} \A \wedge \nabla_{\phi_2} \B \wedge \\
\wedge \left( \sum_{i=1}^k R^{\lambda_i}_{\phantom{\lambda_i}\psi\chi_1 \chi_2} \Clambdai{\psi} - \sum_{i=1}^l R^\psi_{\phantom{\psi}\tau_i \chi_1 \chi_2} \Ctaui{\psi} \right) \\
- R^\psi_{\phantom{\psi}\omega\phi_1 \phi_2} \dx{\omega} \wedge \Bigl( i_\psi \A \wedge \nabla_{\chi_1} \B \wedge \nabla_{\chi_2} \C \\
+ (-1)^{\deg(A)+1} \nabla_{\chi_1} \A \wedge i_\psi \B \wedge \nabla_{\chi_2} \C \\
+ (-1)^{\deg(A)+\deg(B)} \nabla_{\chi_1} \A \wedge \nabla_{\chi_2} \B \wedge i_\psi \C \Bigr) \Biggr] \\
+ \theta^{\phi_1 \chi_1}  \nabla_{\chi_1} \tilde{R}^{\phi_2 \chi_2} \wedge \Bigl( (-1)^{\deg(B)} \nabla_{\phi_1} \A \wedge i_{\phi_2} \B \wedge  i_{\chi_2} \C \\
+ (-1)^{\deg(A)+\deg(B)+1} i_{\chi_2} \A \wedge \nabla_{\phi_1} \B \wedge i_{\phi_2} \C \\
+ (-1)^{\deg(A)} i_{\phi_2} \A \wedge i_{\chi_2} \B \wedge \nabla_{\phi_2} \C \Bigr) \\
+ (-1)^{\deg(B)-1} \left( \tilde{R}^{\phi_1 \chi_1} \wedge  i_{\chi_1} \tilde{R}^{\phi_2 \chi_2} + \tilde{R}^{\phi_2 \chi_1} \wedge  i_{\chi_1} \tilde{R}^{\chi_2 \phi_1} + \tilde{R}^{\chi_2 \chi_1} \wedge  i_{\chi_1} \tilde{R}^{\phi_1 \phi_2} \right) \wedge \\
\wedge i_{\phi_1} \A \wedge i_{\phi_2} \B \wedge  i_{\chi_2} \C \, .\label{Jacobi_graded_nonzero}
\end{multline}
Since the graded Jacobi identity \eqref{Jacobi_graded} requires that the right-hand side of \eqref{Jacobi_graded_nonzero} vanishes, we have to introduce the following constraints:
\begin{enumerate}
\item A covariant version of the Jacobi identity for the Poisson bivector
\begin{equation}
\sum_{(\mu, \nu, \rho)} \theta^{\mu\sigma} \left( \nabla_\sigma \theta^{\nu\rho} - \theta^{\nu\lambda} T^\rho_{\phantom{\rho}\sigma\lambda} \right) = \sum_{(\mu, \nu, \rho)} \theta^{\mu\sigma} \theta^{\nu\lambda} T^\rho_{\phantom{\rho}\sigma\lambda} = 0 \, ,\label{Jacobi_covariant_no2}
\end{equation}
where \eqref{nabla_theta} has been used in the first equality. This constraint is already satisfied \eqref{Jacobi_covariant}.
\item The curvature tensor of the connection $\nabla$ vanishes \eqref{R_zero}.
\item The curvature two-form of the connection $\tilde{\nabla}$ is covariantly constant under $\nabla$,
\begin{equation}
\nabla_\mu \tilde{R}^{\nu\rho} = 0 \, .\label{tildeR_cov_constant_no2}
\end{equation}
This is equivalent to the curvature tensor of $\tilde{\nabla}$ having the same property \eqref{tildeR_cov_constant}.
\item The curvature $\tilde{R}^{\mu\nu}$ satisfies \eqref{RiR}.
\end{enumerate}
\end{description}

Comparing the constraints needed to satisfy the graded differential Poisson algebra of tensor-valued differential forms to the constraints \eqref{tildenabla_symplectic} and \eqref{R_zero}--\eqref{RiR} for differential forms obtained in the literature,  we find that there is no need for new constraints. There are new conditions \eqref{iR_zero}, \eqref{R_T_constraint} and \eqref{Jacobi_covariant_no2} on the connections, but they are all satisfied due to the vanishing of the curvature of the connection $\nabla$ \eqref{R_zero}, the definition of the two connections \eqref{connection_forms} in terms of the same set of connection coefficients and the covariant Jacobi identities \eqref{Jacobi_covariant}.\footnote{See \cite{chu+ho:1997,mccurdy+zumino:2009} for how the condition \eqref{R_T_constraint} is implied by the definition of the two connections \eqref{connection_forms}, the vanishing of the curvature of the connection $\nabla$ \eqref{R_zero} and the so called first Bianchi identity \eqref{DTa} in its tensorial form.}
Thus this generalization to tensor-valued differential forms does not require any additional constraints on the connections.

It has been shown \cite{chu+ho:1997} that due to the constraints \eqref{R_zero} and \eqref{tildenabla_symplectic} there exists a local coordinate system $\{\Phi^\alpha\}$ where the connection coefficients are given in terms of the invertible Poisson bivector $\theta^{\alpha\beta} = \{ \Phi^\alpha, \Phi^\beta \}$ as
\begin{equation}
\Gamma^\alpha_{\beta\gamma} = \theta^{\alpha\delta} \partial_\beta \omega_{\delta\gamma} \, .\label{Gamma}
\end{equation}
Here we refer to these coordinates by the first part of the alphabet $\alpha, \beta, \gamma, \ldots$. The form \eqref{Gamma} of the connection coefficients $\Gamma^\alpha_{\beta\gamma}$ is covariant under the group of affine transformations of the coordinates $\Phi^\alpha$,
\begin{equation}
\Phi^\alpha \rightarrow N^\alpha_{\phantom{\alpha}\beta} \Phi^\beta + V^\alpha \, ,
\end{equation}
where $N^\alpha_{\phantom{\alpha}\beta}$ and $V^\alpha$ are constants, since both sides of \eqref{Gamma} transform like tensors under such affine transformations. The torsion tensor and the (nonvanishing) curvature tensor are, of course, also given by the Poisson structure in these coordinates, e.g. $T^\alpha_{\phantom{\alpha}\beta\gamma} = \theta^{\alpha\delta} \partial_\delta \omega_{\beta\gamma}$. Another special basis is provided by the one-forms $ P_{\alpha\beta} \rd \Phi^\beta$, with respect to which the connection $\nabla$ is trivial, that simplifies many calculations. Most importantly one finds that the Poisson bivector is quadratic in the coordinates $\Phi^\alpha$ by solving the identity $\tilde{R}_{\alpha\beta}^{\phantom{\alpha\beta}\gamma\delta} = \partial_\beta T_\alpha^{\phantom{\alpha}\gamma\delta}$ for the torsion and then the torsion $T_\alpha^{\phantom{\alpha}\beta\gamma} = \partial_\alpha \theta^{\beta\gamma}$ for $\theta^{\alpha\beta}$,\footnote{Here we have used $\theta^{\alpha\beta}$ and $\omega_{\alpha\beta}$ to raise and lower indices respectively. See \cite{chu+ho:1997} for details.}
\begin{equation}
\theta^{\alpha\beta} = \{ \Phi^\alpha, \Phi^\beta \} = \frac{1}{2} \tilde{R}_{\gamma\delta}^{\phantom{\gamma\delta}\alpha\beta} \Phi^\gamma \Phi^\delta + f^{\alpha\beta}_\gamma \Phi^\gamma + g^{\alpha\beta} \, ,\label{theta_quadratic}
\end{equation}
where $\tilde{R}_{\alpha\beta}^{\phantom{\alpha\beta}\gamma\delta}$, $f^{\alpha\beta}_\gamma$ and $g^{\alpha\beta}$ are constants (all antisymmetric under $\alpha \leftrightarrow \beta$). This is somewhat analogous to Darboux's theorem for symplectic geometry.

We provide some further analysis on the constraints imposed on the connections. First we calculate the vanishing covariant derivative $\nabla_\mu$ of $\tilde{R}^{\nu\rho}$ \eqref{tildeR_cov_constant_no2} by using the formula \eqref{nabla_theta} that is implied by the symplecticity of $\tilde{\nabla}$:
\begin{equation}
\nabla_\mu \tilde{R}^{\nu\rho} = \nabla_\mu \left( \theta^{\nu\sigma} \tilde{R}^\rho_{\phantom{\rho}\sigma} \right) = \left( T^\nu_{\phantom{\nu}\mu\lambda} \theta^{\lambda\sigma} + T^\sigma_{\phantom{\sigma}\mu\lambda} \theta^{\nu\lambda} \right) \tilde{R}^\rho_{\phantom{\rho}\sigma} + \theta^{\nu\sigma} \nabla_\mu \tilde{R}^\rho_{\phantom{\rho}\sigma} = 0 \, .
\end{equation}
Multiplying by the symplectic form $\omega_{\tau\nu}$ (sum over $\nu$), introducing the constraint \eqref{R_T_constraint} and renaming some of the indices yields
\begin{equation}
\left( T^\lambda_{\phantom{\lambda}\mu\tau} \omega_{\nu\lambda} \theta^{\tau\sigma} + T^\sigma_{\phantom{\sigma}\mu\nu} \right) \nabla_\sigma T^\rho + \nabla_\mu \nabla_\nu T^\rho = 0 \, .\label{T_constraint}
\end{equation}
Thus the second covariant derivatives of the torsion can be written in terms of first covariant derivatives of the torsion multiplied by the torsion, the Poisson bivector and the symplectic form.

Let us consider the antisymmetric and the symmetric parts of \eqref{T_constraint} with respect to the indices $\mu$ and $\nu$. According to \eqref{cocd_Aform} and the vanishing of the curvature of the connection $\nabla$ \eqref{R_zero} we have $[\nabla_\mu, \nabla_\nu] = - T^\rho_{\phantom{\rho}\mu\nu} \nabla_\rho$. Hence we can decompose
\begin{equation}
\nabla_\mu \nabla_\nu = \nabla_{(\mu} \nabla_{\nu)} + \nabla_{[\mu} \nabla_{\nu]} \\
= \nabla_{(\mu} \nabla_{\nu)} - \frac{1}{2} T^\rho_{\phantom{\rho}\mu\nu} \nabla_\rho \, .\label{nabla2_decomposed}
\end{equation}
Thus the antisymmetric part of \eqref{T_constraint} is
\begin{equation}
\frac{1}{2}\Bigl( \left( T^\lambda_{\phantom{\lambda}\mu\tau} \omega_{\nu\lambda} - T^\lambda_{\phantom{\lambda}\nu\tau} \omega_{\mu\lambda} \right) \theta^{\tau\sigma} + T^\sigma_{\phantom{\sigma}\mu\nu} \Bigr) \nabla_\sigma T^\rho = 0 \, .\label{T_asym_constraint}
\end{equation}
Assuming \eqref{R_T_constraint} does not vanish, \eqref{T_asym_constraint} implies
\begin{equation}
\left( T^\lambda_{\phantom{\lambda}\mu\tau} \omega_{\nu\lambda} - T^\lambda_{\phantom{\lambda}\nu\tau} \omega_{\mu\lambda} \right) \theta^{\tau\sigma} + T^\sigma_{\phantom{\sigma}\mu\nu} = 0
\end{equation}
or equivalently
\begin{equation}
\sum_{(\mu, \nu, \rho)} T^\sigma_{\phantom{\sigma}\mu\nu} \omega_{\sigma\rho} = 0 \, .\label{T_constraint_final}
\end{equation}
Together \eqref{Jacobi_covariant} and \eqref{T_constraint_final} impose a fairly strict set of conditions on the torsion --- though not enough to fix it completely.

The symmetric part of \eqref{T_constraint}, which can be written
\begin{equation}
\nabla_{(\mu} \nabla_{\nu)} T^\rho = \frac{1}{2} \left( T^\sigma_{\phantom{\sigma}\mu\lambda} \omega_{\nu\sigma} + T^\sigma_{\phantom{\sigma}\nu\lambda} \omega_{\mu\sigma} \right) \theta^{\lambda\tau} \nabla_\tau T^\rho \, ,\label{T_sym_constraint}
\end{equation}
does not provide such an interesting result.

\subsection{Star product}
\label{sec:star_product}
The star product for tensor-valued differential forms can be defined similarly as in \cite{mccurdy+zumino:2009}
\begin{equation}
\begin{split}
\A \star \B &= \A \wedge \B \\
&+ \sum_{n=1}^\infty \hbar^n C_n \left( \A, \B \right) \, ,\label{star_product}
\end{split}
\end{equation}
where $C_n$ are bilinear covariant differential operators of at most order $n$ in each argument, which are constructed from the covariant derivatives $\nabla_\mu$, the Poisson bivector $\theta$, the torsion tensor and the curvature tensor(s). Further the operators $C_n$ are chosen so that the star product \eqref{star_product} satisfies the following properties:
\begin{enumerate}
\item The star product is associative
\begin{multline}
\A \star \bigl( \B \star \C \bigr) \\
= \left( \A \star \B \right) \star \C \, .\label{associativity}
\end{multline}
\item The first order deformation is given by the Poisson bracket \eqref{Pb_tensor-valued}
\begin{equation}
C_1 \left( \A, \B \right) = \left\{ \A, \B \right\} \, .
\end{equation}
\item The constant function, $M \ni x \mapsto 1$, is the identity: $1 \star A = A \star 1 = A$.
\item Every $C_n$ is of order $n$ in the Poisson bivector $\theta$ (including its covariant derivatives \eqref{nabla_theta} and the curvature \eqref{tildeR}) and it has the degree
\begin{equation}
\deg\Bigl( C_n \left( \A, \B \right) \Bigr) = \deg(A) + \deg(B) \, .
\end{equation}
\item The operators $C_n$ have the generalized Moyal symmetry
\begin{equation}
C_n \left( \A, \B \right) = (-1)^{\deg(A) \deg(B) + n} C_n \left( \B, \A \right) \, .\label{generalized_Moyal_symmetry}
\end{equation}
\end{enumerate}

To the second order in the deformation parameter $\hbar$ the star product is given by
\begin{multline}
C_2 \left( \A, \B \right) = \frac{1}{2} \theta^{\lambda_1 \tau_1} \theta^{\lambda_2 \tau_2} \nabla_{\lambda_1} \nabla_{\lambda_2} \A \wedge \nabla_{\tau_1} \nabla_{\tau_2} \B \\
+ \frac{1}{3} \left( \theta^{\lambda_1 \tau_1} \nabla_{\tau_1} \theta^{\lambda_2 \tau_2} + \frac{1}{2} \theta^{\lambda_2 \phi} \theta^{\tau_2 \chi} T^{\lambda_1}_{\phantom{\lambda_1} \phi\chi} \right) \Bigl( \nabla_{\lambda_1} \nabla_{\lambda_2} \A \wedge \nabla_{\tau_2} \B \\
+ \nabla_{\tau_2} \A \wedge \nabla_{\lambda_1} \nabla_{\lambda_2} \B \Bigr) \\
+ (-1)^{\deg(A)} \theta^{\lambda_1 \tau_1} \tilde{R}^{\lambda_2 \tau_2} \wedge \nabla_{\lambda_1} i_{\lambda_2} \A \wedge \nabla_{\tau_1} i_{\tau_2} \B \\
- \frac{1}{2} \tilde{R}^{\lambda_1 \tau_1} \wedge \tilde{R}^{\lambda_2 \tau_2} \wedge i_{\lambda_1} i_{\lambda_2} \A \wedge i_{\tau_1} i_{\tau_2} \B \\
- \frac{1}{3} \tilde{R}^{\lambda_1 \tau_1} \wedge i_{\tau_1} \tilde{R}^{\lambda_2 \tau_2} \wedge \Bigl( (-1)^{\deg(A)} i_{\lambda_1} i_{\lambda_2} \A \wedge i_{\tau_2} \B \\
+ i_{\lambda_2} \A \wedge i_{\lambda_1} i_{\tau_2} \B \Bigr) \, .\label{C_2_tensor-valued}
\end{multline}
The second term of \eqref{C_2_tensor-valued} can be simplified by using \eqref{nabla_theta} and \eqref{Jacobi_covariant},
\begin{equation}
\theta^{\mu\sigma} \nabla_\sigma \theta^{\nu\rho} + \frac{1}{2} \theta^{\nu\sigma} \theta^{\rho\lambda} T^\mu_{\phantom{\mu}\sigma\lambda} = - \frac{1}{2} \theta^{\nu\sigma} \theta^{\rho\lambda} T^\mu_{\phantom{\mu}\sigma\lambda} \, ,
\end{equation}
but we choose to keep the similarity with the star product of \cite{mccurdy+zumino:2009}.\footnote{There is a sign difference in the second factor of the second term of \eqref{C_2_tensor-valued} compared to \cite{mccurdy+zumino:2009} that is enabled by the antisymmetry of the first factor under $\lambda_2 \leftrightarrow \tau_2$. The motivation for this cosmetic change is to emphasize the symmetry property \eqref{generalized_Moyal_symmetry} of $C_2$.}
Proof of the associativity of the star product \eqref{star_product} to $\mc{O}(\hbar^2)$ is completely analogous with \cite{mccurdy+zumino:2009}.\footnote{Due to the vanishing of the curvature of the connection $\nabla$ the tensorial indices can mostly be ignored in the calculation verifying the associativity \eqref{associativity} to $\mc{O}(\hbar^2)$.}
At the classical level $\mc{O}(1)$ the associativity is trivially implied by the associativity of the exterior product. At $\mc{O}(\hbar)$ the associativity is implied by the graded symmetry rule \eqref{graded_product}. At $\mc{O}(\hbar^2)$ the associativity condition
\begin{equation}
\begin{split}
&\A \wedge C_2 \left( \B, \C \right) \\
&- C_2 \left( \A \wedge \B, \C \right)\\
&+ C_2 \left( \A, \B \wedge \C \right) \\
&- C_2 \left( \A, \B \right) \wedge \C \\
&= C_1 \Bigl( C_1 \left( \A, \B \right), \C \Bigr) \\
&- C_1 \Bigl( \A, C_1 \left( \B, \C \right) \Bigr)
\label{associativity_2nd_order}
\end{split}
\end{equation}
can be shown to hold by using the properties of the Poisson bracket, the constraints these properties imply --- namely \eqref{tildenabla_symplectic}, \eqref{Jacobi_covariant}, \eqref{R_zero}, \eqref{tildeR_cov_constant} and \eqref{RiR} --- and the properties of the covariant derivative and the interior product --- including the commutativity of the two \eqref{i_and_nabla_commute}, $i_\mu i_\nu = - i_\nu i_\mu$ and the decomposition \eqref{nabla2_decomposed}.

As discussed in \cite{mccurdy+zumino:2009} the next order $\hbar^3$ deformation could be derived with a considerable amount of calculation by finding an ansatz that satisfies the required conditions.

If the torsion vanishes, we have a flat symplectic connection $\nabla$. Then the star product \eqref{star_product} can be defined by 
\begin{multline}
\left. \A \star \B \right|_{T=0} = \A \wedge \B \\
+ \sum_{n=1}^\infty \frac{\hbar^n}{n!}\theta^{\lambda_1 \tau_1} \cdots \theta^{\lambda_n \tau_n} \nabla_{\lambda_1} \cdots \nabla_{\lambda_n} \A \wedge \nabla_{\tau_1} \cdots \nabla_{\tau_n} \B \, ,
\end{multline}
since now the covariant derivatives commute both with each other and with the Poisson bivector $\theta^{\mu\nu}$.

\subsection{On the algebra of tensors}
Thus by starting from the graded differential Poisson structure on the algebra of forms $\Omega(M)$, we have generalized it to the algebra of tensor-valued differential forms \eqref{algebra_tvf} and consequently to the subalgebra of all tensor fields on $M$,
\begin{equation}
\mc{T}(M) = \bigoplus_{k,l=0}^\infty \Omega^0(M, T^{k, l}) \subset \Omega(M, T) \, .
\end{equation}
For such tensor-valued zero-forms the Poisson bracket \eqref{Pb_tensor-valued} is reduced to \eqref{Pb_tensor} and in the star product,
\begin{equation}
\begin{split}
\A \star \B &= \A \B \\
&+ \sum_{n=1}^\infty \hbar^n C_n \left( \A, \B \right) \, ,\label{star_product_tensor}
\end{split}
\end{equation}
the deformation of order $\hbar^2$ is written
\begin{multline}
C_2 \left( \A, \B \right) = \frac{1}{2} \theta^{\lambda_1 \tau_1} \theta^{\lambda_2 \tau_2} \nabla_{\lambda_1} \nabla_{\lambda_2} \A \nabla_{\tau_1} \nabla_{\tau_2} \B \\
+\frac{1}{3} \left( \theta^{\lambda_1 \tau_1} \nabla_{\tau_1} \theta^{\lambda_2 \tau_2} + \frac{1}{2} \theta^{\lambda_2 \phi} \theta^{\tau_2 \chi} T^{\lambda_1}_{\phantom{\lambda_1} \phi\chi} \right) \Bigl( \nabla_{\lambda_1} \nabla_{\lambda_2} \A \nabla_{\tau_2} \B \\
+ \nabla_{\tau_2} \A \nabla_{\lambda_1} \nabla_{\lambda_2} \B \Bigr) \, .\label{C_2_tensor}
\end{multline}

In the case of vanishing torsion we obtain the simple star product of tensor fields
\begin{equation}
\left. \A \star \B \right|_{T=0} = \A \exp\left( \hbar \overleftarrow{\nabla}_\lambda \theta^{\lambda\tau} \overrightarrow{\nabla}_\tau \right) \B \, .\label{reduced_star_product_tensor}
\end{equation}
In the recent work \cite{vassilevich:2009} a covariant star product of functions was defined on a symplectic manifold with vanishing torsion and curvature ($T=R=0$). It was also proposed that this star product could be straightforwardly extented for tensor fields. We recognize that the reduced ($T=R=\tilde{R}=0$) case \eqref{reduced_star_product_tensor} of our more general star product of tensor fields \eqref{star_product_tensor} is exactly what the extension of the star product of \cite{vassilevich:2009} to tensor fields would be.

\subsection{Discussion}
When we consider possible applications of these star products \eqref{star_product} and \eqref{star_product_tensor} in physics, particularly gravity and gauge theory, the problem (perhaps also a possibility) is that the structure of the graded differential Poisson algebra of (tensor-valued) differential forms requires strict constraints on the underlying symplectic manifold. Due to the required constraints the torsion and the curvature are rather restricted, which is likely to cause some problems particularly for theories of gravity. Still the connection $\nabla$ can have a nonvanishing torsion and in this case the symplectic connection $\tilde{\nabla}$ also has curvature. This should open up the possibility for some nontrivial gravitational dynamics.

 In the extremely restricted ($T=R=\tilde{R}=0$) case \eqref{reduced_star_product_tensor} that was also recently studied in \cite{vassilevich:2009} there is virtually impossible to have a nontrivial theory of gravity, because neither the energy-momentum tensor nor the spin density tensor are supported due to the vanishing of both the curvature and the torsion. Setting up the equivalence principle would clearly be impossible. Thus this star product \eqref{reduced_star_product_tensor} can be used only in cases where the curvature and the torsion vanish in the corresponding commutative theory. Then in the noncommutative extension of the theory we would find corrections to the geometrical objects in the higher orders of the deformation parameter $\hbar$ due to the star product. In the gravitational field equations these corrections would require compensating corrections to the energy-momentum tensor and possibly to the spin density tensor depending on the chosen action. This is problematic since, as we noted, matter fields are not supported in this case.\footnote{Such corrections to the right hand (energy-momentum) side of field equations frequently appear in noncommutative theories of gravity when a star product is introduced. Particularly in the case of vacuum field equations such corrections cannot be associated to matter fields, because presumably there is no matter in empty space. So the corrections would have to be physically interpreted as some kind of energy-momentum inherent to the noncommutative spacetime. However, at this point such interpretations are mere speculations.}
An example of such theory is the two-dimensional noncommutative dilaton gravity studied in \cite{vassilevich:2009}.

In the case of gauge theory these restrictions are not quite as severe as in the case of gravity. Noncommutative gauge theory with Yang-Mills actions has been studied \cite{ho+miao:2001,chaichian+tureanu+zet:2009,chaichian+oksanen+tureanu+zet:2010} in this setting. The former work employed the popular Seiberg-Witten map \cite{seiberg+witten:1999}. In \cite{chaichian+tureanu+zet:2009,chaichian+oksanen+tureanu+zet:2010} the star product of differential forms was generalized to Lie algebra-valued differential forms in order to be able to apply it to the connection one-form of the gauge theory, as well as to the gauge transformation parameter and the field strength, which are all Lie algebra-valued. This generalization is fairly simple to achieve since the generators of the internal gauge symmetry commute with the covariant derivation $\nabla$. The generalization to tensor-valued differential forms we have presented can be further generalized to Lie algebra-valued objects,
\begin{equation}
\A = \ALiea T_a \, ,
\end{equation}
where $T_a$ are the generators of the Lie algebra, along the lines of \cite{chaichian+tureanu+zet:2009} with relative ease.\footnote{Please note that one of the misunderstandings of \cite{tagliaferro:2008} has been inherited to \cite{chaichian+tureanu+zet:2009}. Namely, $[\nabla_\mu, \nabla_\nu] \alpha = 0$ is not required for any $\alpha \in \Omega(M)$, since it would also imply that the torsion vanishes, which is not necessary.}
The star product is defined by
\begin{equation}
\begin{split}
\A \star \B &= \ALiea \wedge \BLieb T_a T_b \\
&+ \sum_{n=1}^\infty \hbar^n C_n \left( \ALiea, \BLieb \right) T_a T_b \, ,
\end{split}
\end{equation}
where the operators $C_n$ are defined as before. In order to obtain a star commutator that is consistent with \cite{chaichian+tureanu+zet:2009,chaichian+oksanen+tureanu+zet:2010} we have required the symmetry property \eqref{generalized_Moyal_symmetry} for $C_n$, though it is not required in \cite{mccurdy+zumino:2009}.
\begin{equation}
\begin{split}
\left[ \A, \B \right]_\star &\equiv \A \star \B \\
&- (-1)^{\deg(A) \deg(B)} \B \star \A \\
&= \ALiea \wedge \BLieb [T_a, T_b] \\
&+ \sum_{n=1}^\infty \hbar^n C_n \left( \ALiea, \BLieb \right) [T_a, T_b]_{(n)} \, ,
\end{split}
\end{equation}
where $[T_a, T_b]_{(n)} = T_a T_b - (-1)^n T_b T_a$ is the anticommutator, $\{T_a, T_b\}$, for every odd $n$ and the commutator, $[T_a, T_b]$, for every even $n$.

\section{Covariant star product on a Poisson manifold}
In this section we discuss a covariant star product on a regular Poisson manifold $M$, first for functions and then for tensor fields. Since $M$ is regular, we can require that a linear connection exists on $M$. On a nonregular Poisson manifold we would generally define a different connection on each symplectic leaf of $M$, or define a contravariant connection on $M$ and use the associated contravariant derivative instead of a covariant one \cite{vaisman:1991,book:vaisman:1994,fernandes:2000}.

It was shown by Kontsevich \cite{kontsevich:2003} that a star product can be constructed for smooth functions on $\bb{R}^d$ with any Poisson structure $\theta$ in the sense of deformation quantization, so that at first order in the deformation parameter the star product is given by the Poisson bracket of functions. A path integral formulation of the Kontsevich quantization has been developed \cite{cattaneo+felder:2000}. The Kontsevich formula is not well-suited for calculating the star product beyond $\hbar^2$, because it contains integrals that cannot be solved by any standard method. The star product of functions has been calculated up to $\hbar^4$ by using a simpler iterative approach \cite{kupriyanov+vassilevich:2008}.

The existence of a covariant star product of functions on any Poisson manifold $(M, \theta)$ with a torsion-free linear connection has been shown in \cite{ammar+chloup+gutt:2008} and given explicitly to $\hbar^3$ as an example.

\subsection{Star product of functions}
The Poisson structure on the algebra of smooth functions $f, g \in \mc{F}(M)$ is defined by
\begin{equation}
\{ f, g \} = \theta( \rd f, \rd g ) = \theta^{\mu\nu} \partial_\mu f \partial_\nu g \, ,\label{Pb_function}
\end{equation}
where $\theta$ is a Poisson bivector field, i.e. a smooth section of $\wedge^2 TM$. The Poisson bracket \eqref{Pb_function} satisfies the required properties
\begin{enumerate}
\item Antisymmetry: $\{ f, g \} = - \{ g, f \}$
\item Jacobi identity:
\begin{equation}
\{ f, \{ g, h \} \} + \{ g, \{ h, f\} \} + \{ h, \{ f, g\} \} = 0
\end{equation}
\item Derivation in the second argument:
\begin{equation}
\{ f, gh \} = \{ f, g \} h + g \{ f, h \}
\label{derivation_2nd_arg}
\end{equation}
\end{enumerate}
 when the bivector $\theta^{\mu\nu}$ satisfies the Jacobi identity \eqref{Jacobi}.

The star product of functions $f, g \in \mc{F}(M)$ is defined by
\begin{equation}
f \star g = fg + \sum_{n=1}^\infty \hbar^n C_n (f, g) \, ,\label{star_product_function}
\end{equation}
where the bidifferential operators $C_n : \mc{F}(M) \times \mc{F}(M) \rightarrow \mc{F}(M)$ are constructed from the torsion-free linear connection $\nabla$, the Poisson bivector and the curvature tensor. At order $\hbar$ one has
\begin{equation}
C_1 (f, g) = \{ f, g \} \, .
\end{equation}
The star product \eqref{star_product_function} is required to be associative to all orders in $\hbar$,
\begin{equation}
f \star (g \star h) = (f \star g) \star h \, .
\end{equation}
Such star product of functions is given to order $\hbar^3$ by \cite{ammar+chloup+gutt:2008}
\begin{align}
C_2 (f, g) &= \frac{1}{2} \theta^{\mu\nu}\theta^{\rho\sigma} \nabla_\mu \nabla_\rho f \nabla_\nu \nabla_\sigma g + \frac{1}{3} \theta^{\mu\sigma} \nabla_\sigma \theta^{\nu\rho} ( \nabla_\mu \nabla_\nu f \nabla_\rho g + \nabla_\rho f \nabla_\mu \nabla_\nu g ) \label{C_2_function}\\
&+ \frac{1}{6} \nabla_\rho \theta^{\mu\nu} \nabla_\mu \theta^{\rho\sigma} \nabla_\nu f \nabla_\sigma g \, ,\nonumber\\
C_3 (f, g) &= - \frac{1}{6} \theta^{\rho\sigma} \left( \mc{L}_{X_f} \nabla \right)^\mu_{\nu\rho} \left( \mc{L}_{X_g} \nabla \right)^\nu_{\mu\sigma} \, ,
\end{align}
where $\mc{L}_{X_f} \nabla$ is the tensor defined by the Lie derivative of the connection $\nabla$ along the Hamiltonian vector field $X_f = i(\rd f) \theta$:
\begin{equation}
\begin{split}
\left( \mc{L}_{X_f} \nabla \right)^\mu_{\nu\rho} &= \theta^{\sigma\mu} \nabla_\nu \nabla_\rho \nabla_\sigma f + \nabla_\nu \theta^{\sigma\mu} \nabla_\rho \nabla_\sigma f + \nabla_\rho \theta^{\sigma\mu} \nabla_\nu \nabla_\sigma f \\
&+ \nabla_\nu \nabla_\rho \theta^{\sigma\mu} \nabla_\sigma f + R^\mu_{\phantom{\mu}\sigma\nu\rho} \theta^{\sigma\lambda} \nabla_\lambda f \, .\label{L_X_nabla}
\end{split}
\end{equation}
Note that since the torsion vanishes, $T^\rho_{\phantom{\rho}\mu\nu} = 0$, the covariant derivatives commute
\begin{equation}
[\nabla_\mu, \nabla_\nu] f = - T^\rho_{\phantom{\rho}\mu\nu} \nabla_\rho f = 0 \, ,\label{cocd_function_zero}
\end{equation}
for every $f \in \mc{F}(M)$. This star product exists for any Poisson manifold $(M, \theta)$ and any torsion-free connection $\nabla$.

A covariant star product of functions can alternatively be defined directly according to the Kontsevich universal formula \cite{kontsevich:2003} by replacing the partial derivatives $\partial_\mu$ with the covariant derivatives $\nabla_\mu$ in all $C_n$, $n > 1$. By using the results of \cite{kupriyanov+vassilevich:2008} one can write this star product up to order $\hbar^4$. At orders higher than $\hbar^2$, where second and higher covariant derivatives of the bivector $\theta$ appear, we have to introduce another condition
\begin{equation}
[\nabla_\mu, \nabla_\nu] \theta^{\rho\sigma} = 0 \, ,\label{cocd_theta}
\end{equation}
in addition to the vanishing of the torsion \eqref{cocd_function_zero}, in order to ensure the associativity of the star product. Thus the curvature tensor of the connection satisfies
\begin{equation}
\begin{split}
[\nabla_\mu, \nabla_\nu] \theta^{\rho\sigma} &= -T^\lambda_{\phantom{\lambda}\mu\nu} \nabla_\lambda  \theta^{\rho\sigma} + R^\rho_{\phantom{\rho}\lambda\mu\nu} \theta^{\lambda\sigma} + R^\sigma_{\phantom{\sigma}\lambda\mu\nu} \theta^{\rho\lambda} \\
&= - \theta^{\sigma\lambda} R^\rho_{\phantom{\rho}\lambda\mu\nu} + \theta^{\rho\lambda} R^\sigma_{\phantom{\sigma}\lambda\mu\nu} = 0
\end{split}
\end{equation}
or equivalently
\begin{equation}
R^{\mu\nu}_{\phantom{\mu\nu}\rho\sigma} = R^{\nu\mu}_{\phantom{\nu\mu}\rho\sigma} \, .
\end{equation}
It is sufficient for $\theta$ to be covariantly constant, $\nabla_\mu \theta^{\nu\rho} = 0$, but it is not necessary.\footnote{In the case of a symplectic manifold the condition $\nabla_\mu \theta^{\nu\rho} = 0$ would be equivalent to the symplectic form $\omega$ to be covariantly constant, i.e. having a symplectic torsion-free connection.}
Without the above condition for the curvature we would have to add terms with curvature contributions to the star product in order to satisfy the associativity requirement. Indeed this approach is nothing more than a special case of the universal star product studied in \cite{ammar+chloup+gutt:2008}.

Relaxing the condition of \cite{ammar+chloup+gutt:2008} that the connection is torsion-free appears to be very difficult without imposing some constraints on both the curvature and the torsion. We shall discuss this briefly while considering a star product of tensor fields.

\subsection{Star product of tensor fields}
Although we have found a covariant star product of tensor fields on a symplectic manifold as a special case of a star product of tensor-valued differential forms in the Section~\ref{sec:tensor-valued_forms}, we would like a find a construction with less constraints on the connection. Since it is the definition of the Poisson bracket that primarily imposes the constraints on the connections in the case of tensor-valued differential forms, we attempt to define a Poisson bracket of tensor fields with a minimal set of properties.

The Poisson structure \eqref{Pb_function} can be extended on the algebra of smooth tensor fields $A, B \in \mc{T}(M)$ by
\begin{equation}
\left\{ \A, \B \right\} = \theta^{\lambda\tau} \nabla_\lambda \A \nabla_\tau \B \, .\label{Pb_tensor_again}
\end{equation}
For a function $f \in \mc{F}(M)$ the bracket $\{f, \cdot\}$ is a covariant derivation with respect to the second argument
\begin{equation}
\left\{ f, \A \right\} = \nabla_{X_f} \A \, ,
\end{equation}
with the Hamiltonian vector field $X _f^\mu = \theta^{\nu\mu} \nabla_\nu f$. We postulate the following properties for the Poisson bracket as a straightforward generalization of the usual case of functions.
\begin{enumerate}
\item Antisymmetry:
\begin{equation}
\left\{ \A, \B \right\} = - \left\{ \B, \A \right\}
\end{equation}
\item Jacobi identity:
\begin{equation}
\begin{split}
&\left\{ \A, \bigl\{ \B, \C \bigr\} \right\} \\
&+ \left\{ \B, \bigl\{ \C, \A \bigr\} \right\} \\
&+ \left\{ \C, \left\{ \A, \B \right\} \right\} = 0 \label{Jacobi_identity_tensor}
\end{split}
\end{equation}
\item ``Derivation in the second argument'':
\begin{equation}
\begin{split}
\bigl\{ \A, \B \C \bigr\} &= \left\{ \A, \B \right\} \C \\
&+ \B \bigl\{ \A, \C \bigr\} \label{Pb2_ABC}
\end{split}
\end{equation}
\end{enumerate}
The Jacobi identity \eqref{Jacobi_identity_tensor} imposes the following two constraints on the connection $\nabla$:
\begin{gather}
\sum_{(\mu, \nu, \rho)} \theta^{\mu\sigma} \left( \nabla_\sigma \theta^{\nu\rho} - \theta^{\nu\lambda} T^\rho_{\phantom{\rho}\sigma\lambda} \right) = 0 \, ,\label{Jacobi_covariant_no3}\\
\theta^{\mu\lambda} \theta^{\nu\tau} R^\rho_{\phantom{\rho}\sigma\lambda\tau} = 0 \, .\label{theta2_R_zero}
\end{gather}
Note that according to \eqref{theta2_R_zero} the curvature tensor does not need to vanish everywhere since the Poisson bivector $\theta^{\mu\nu}$ is not necessarily invertible.

Then we quantize the Poisson manifold by defining a covariant star product of tensor fields as in \eqref{star_product_tensor}. The order $\hbar$ deformation, $C_1$, is again defined to be the Poisson bracket \eqref{Pb_tensor_again}. The operators $C_n$ are chosen to satisfy the same properties as in the Section~\eqref{sec:star_product}.\footnote{The sign factor in the symmetry property \eqref{generalized_Moyal_symmetry} is obviously replaced with $(-1)^n$.}
A propriate ansatz for $C_2$ can be found by calculating the side of the associativity condition at order $\hbar^2$ that depends on $C_1$ and choosing a $C_2$ that produces a similar expression on the other side of the condition. We choose $C_2$ to be of the same form as in \eqref{C_2_tensor}. The associativity property of the star product imposes the additional constraint
\begin{equation}
\sum_{(\mu, \nu, \rho)} \theta^{\mu\sigma} \left( \nabla_\sigma \theta^{\nu\rho} + \frac{1}{2} \theta^{\nu\lambda} T^\rho_{\phantom{\rho}\sigma\lambda} \right) = 0 \, .\label{Jacobi_covariant_no4}
\end{equation}
In order to satisfy both \eqref{Jacobi_covariant_no3} and \eqref{Jacobi_covariant_no4} we require that the connection satisfies the covariant Jacobi identities \eqref{Jacobi_covariant}, so that the cyclic sum over each of the terms of \eqref{Jacobi_covariant_no3} and \eqref{Jacobi_covariant_no4} is zero. Thus the constraints \eqref{Jacobi_covariant} and \eqref{theta2_R_zero} are all that is needed for a covariant star product of tensor fields on a Poisson manifold up to order $\hbar^2$.

In the case of a star product of functions there is no need for the constraint \eqref{theta2_R_zero}. However, the other constraints \eqref{Jacobi_covariant} are required, and they constrain the connection so that both the torsion and the curvature are affected. Thus relaxing the torsion-freeness constraint has lead to having some constraints for both the torsion and the curvature.

At present it is unclear whether additional constraints need to be introduced for the connection at higher orders in $\hbar$.

\section{Conclusion}
We have generalized the recently defined covariant star product of differential forms on a symplectic manifold \cite{mccurdy+zumino:2009} to tensor-valued differential forms and consequently to tensor fields of any type. This generalization does not require any new constraints on the connections. Possible applications of the star product to gravity and gauge theory have been discussed, considering the rather strict constraints the connections have to satisfy. Further study of both of these applications is required.

Then we proposed a covariant star product of tensor fields on a Poisson manifold with a linear connection that has less constraints than in the first case. Thus this star product could be a more viable option for theories of gravity.

We also discussed the possibility to relax the torsion-freeness condition of the linear connection of the universal covariant star product of functions defined on a Poisson manifold in \cite{ammar+chloup+gutt:2008}. It was found that this requires one to impose some constraints on both the torsion and the curvature, namely \eqref{Jacobi_covariant} in our case.

Finally, a remark about the Poisson algebra of tensor fields is in order. A Poisson algebra consists of a commutative associative algebra endowed with a Poisson bracket. A graded-commutative associative algebra --- like the algebra of differential forms --- can be turned into a graded Poisson algebra by introducing a graded Poisson bracket. However, the algebra of tensor fields is neither commutative nor graded-commutative. This is the reason why the Poisson structure of tensor fields \eqref{Pb_tensor_again}---\eqref{Pb2_ABC} was defined for the components of the tensors, which are of course commutative functions. This raises the question could a Poisson structure for tensor fields be defined some other way compared to the definition given above?

\section*{Acknowledgments}
The support of the Academy of Finland under the Projects No. 121720 and 127626 is greatly acknowledged. The work of M. O. was fully supported by the Jenny and Antti Wihuri Foundation. G. Z. acknowledges the support of CNCSIS-UEFISCSU Grant ID-620 of the Ministry of Education and Research of Romania.


\begin{thebibliography}{99}
\addcontentsline{toc}{section}{References}
\providecommand{\href}[2]{#2}

\bibitem{douglas+nekrasov:2001}
M.~R. Douglas and N.~A. Nekrasov, Noncommutative field theory,
  \href{http://dx.doi.org/10.1103/RevModPhys.73.977}{{\em Rev. Mod. Phys.} {\bf
  73} (2001)  977}, \href{http://arxiv.org/abs/hep-th/0106048}{{\tt
  arXiv:hep-th/0106048}}.

\bibitem{szabo:2003}
R.~J. Szabo, Quantum field theory on noncommutative spaces,
  \href{http://dx.doi.org/10.1016/S0370-1573(03)00059-0}{{\em Phys. Rept.} {\bf
  378} (2003)  207}, \href{http://arxiv.org/abs/hep-th/0109162}{{\tt
  arXiv:hep-th/0109162}}.

\bibitem{mccurdy+zumino:2009}
S.~McCurdy and B.~Zumino, Covariant star product for exterior
  differential forms on symplectic manifolds,
  \href{http://dx.doi.org/10.1063/1.3327559}{{\em AIP Conf. Proc.} {\bf 1200} (2010) 204},
  \href{http://arxiv.org/abs/0910.0459}{{\tt arXiv:0910.0459 [hep-th]}}.

\bibitem{tagliaferro:2008}
A.~Tagliaferro, A star product for differential forms on symplectic manifolds, \href{http://arxiv.org/abs/0809.4717}{{\tt arXiv:0809.4717
  [hep-th]}} (2008).

\bibitem{vassilevich:2009}
D.~V. Vassilevich, Diffeomorphism covariant star products and noncommutative
  gravity, \href{http://dx.doi.org/10.1088/0264-9381/26/14/145010}{{\em Class.
  Quant. Grav.} {\bf 26} (2009) 145010} , \href{http://arxiv.org/abs/0904.3079}{{\tt
  arXiv:0904.3079 [hep-th]}}.

\bibitem{ammar+chloup+gutt:2008}
M.~Ammar, V.~Chloup, and S.~Gutt, Universal star products,
  \href{http://dx.doi.org/10.1007/s11005-008-0240-0}{{\em Lett. Math. Phys}
  {\bf 84} (2008)  199}, \href{http://arxiv.org/abs/0804.1300}{{\tt
  arXiv:0804.1300 [math.SG]}}.

\bibitem{dito+stemheimer:2002}
G.~Dito and D.~Sternheimer, Deformation quantization: genesis, developments and
  metamorphoses, in {\em IRMA Lectures in Math. Theoret. Phys.}, W.~D. Gruyter,
  ed., vol.~1, p.~9.
\newblock 2002.
\newblock \href{http://arxiv.org/abs/math/0201168}{{\tt arXiv:math/0201168}}.

\bibitem{book:kobayashi+nomizu:1963}
S.~Kobayashi and K.~Nomizu, {\em Foundations of differential geometry}, vol.~1.
\newblock Interscience Publishers, New York and London, 1963.

\bibitem{chu+ho:1997}
C.-S. Chu and P.-M. Ho, Poisson algebra of differential forms,
  \href{http://dx.doi.org/10.1142/S0217751X97002929}{{\em Int. J. Mod. Phys.}
  {\bf A 12} (1997)  5573}, \href{http://arxiv.org/abs/q-alg/9612031}{{\tt
  arXiv:q-alg/9612031}}.

\bibitem{beggs+majid:2006}
E.~J. Beggs and S.~Majid, Semi-classical differential structures, {\em Pacific
  Journal of Mathematics} {\bf 224} (2006)  1,
  \href{http://arxiv.org/abs/math/0306273}{{\tt arXiv:math/0306273}}.

\bibitem{ho+miao:2001}
P.-M. Ho and S.-P. Miao, Noncommutative differential calculus for {D}-brane in
  non-constant {B} field background with {H=0},
  \href{http://dx.doi.org/10.1103/PhysRevD.64.126002}{{\em Phys. Rev.} {\bf D
  64} (2001)  126002}, \href{http://arxiv.org/abs/hep-th/0105191}{{\tt
  arXiv:hep-th/0105191}}.

\bibitem{chaichian+tureanu+zet:2009}
M.~Chaichian, A.~Tureanu, and G.~Zet, Gauge field theories with covariant
  star-product, \href{http://dx.doi.org/10.1088/1126-6708/2009/07/084}{{\em
  JHEP} {\bf 07} (2009)  084}, \href{http://arxiv.org/abs/0905.0608}{{\tt
  arXiv:0905.0608 [hep-th]}}.

\bibitem{chaichian+oksanen+tureanu+zet:2010}
M.~Chaichian, M.~Oksanen, A.~Tureanu, and G.~Zet, 
  Noncommutative gauge theory using a covariant star product defined between Lie-valued differential forms,
  \href{http://dx.doi.org/10.1103/PhysRevD.81.085026}{{\em
 Phys. Rev.} {\bf D 81} (2010)  085026}, 
  \href{http://arxiv.org/abs/1001.0508}{{\tt arXiv:1001.0508 [hep-th]}}.

\bibitem{seiberg+witten:1999}
N.~Seiberg and E.~Witten, String theory and noncommutative geometry, {\em JHEP}
  {\bf 09} (1999)  032, \href{http://arxiv.org/abs/hep-th/9908142}{{\tt
  arXiv:hep-th/9908142}}.

\bibitem{vaisman:1991}
I.~Vaisman, On the geometric quantization of poisson manifolds,
  \href{http://dx.doi.org/10.1063/1.529446}{{\em J. Math. Phys.} {\bf 32}
  (1991)  3339}.

\bibitem{book:vaisman:1994}
I.~Vaisman, {\em Lectures on the geometry of Poisson manifolds}, vol.~118 of
  {\em Progress in mathematics}.
\newblock Birkh\"auser, Berlin, 1994.

\bibitem{fernandes:2000}
R.~L. Fernandes, Connections in {P}oisson geometry {I}: Holonomy and
  invariants, {\em J. Diff. Geom.} {\bf 54} (2000)  303,
  \href{http://arxiv.org/abs/math/0001129}{{\tt arXiv:math/0001129}}.

\bibitem{kontsevich:2003}
M.~Kontsevich, Deformation quantization of {P}oisson manifolds,
  \href{http://dx.doi.org/10.1023/B:MATH.0000027508.00421.bf}{{\em Lett. Math.
  Phys} {\bf 66} (2003)  157}, \href{http://arxiv.org/abs/q-alg/9709040}{{\tt
  arXiv:q-alg/9709040}}.

\bibitem{cattaneo+felder:2000}
A.~S. Cattaneo and G.~Felder, A path integral approach to the {K}ontsevich
  quantization formula, {\em Commun. Math. Phys} {\bf 212} (2000)  591,
  \href{http://arxiv.org/abs/math/9902090}{{\tt arXiv:math/9902090}}.

\bibitem{kupriyanov+vassilevich:2008}
V.~G. Kupriyanov and D.~V. Vassilevich, Star products made (somewhat) easier,
  \href{http://dx.doi.org/10.1140/epjc/s10052-008-0804-2}{{\em Eur. Phys. J.}
  {\bf C 58} (2008)  627}, \href{http://arxiv.org/abs/0806.4615}{{\tt
  arXiv:0806.4615 [hep-th]}}.

\end{thebibliography}
\end{document}